\documentclass[twocolumn]{aastex631}

\usepackage{graphicx}
\usepackage{amsmath}
\usepackage{multirow}
\usepackage{float}
\usepackage{verbatim}

\usepackage{hyperref}
\usepackage{color}

\usepackage[normalem]{ulem}
\usepackage{txfonts}
\usepackage[marginal]{footmisc}

\usepackage{bm}

\begin{document}

\title{X-ray spectral correlations in a sample of Low-mass black hole X-ray binaries in the hard state}

\correspondingauthor{Bei You, Youli Tuo, Xinwu Cao}
\email{youbei@whu.edu.cn, tuo@astro.uni-tuebingen.de, xwcao@zju.edu.cn}

\author{Bei You}
\affiliation{School of Physics and Technology, Wuhan University, Wuhan, 430072, People’s Republic of China}

\author{Yanting Dong}
\affiliation{Institute for Astronomy, School of Physics, Zhejiang University, 866 Yuhangtang Road, Hangzhou, 310058, People’s Republic of China}

\author{Zhen Yan}
\affiliation{Shanghai Astronomical Observatory, Chinese Academy of Sciences, 80 Nandan Road, Shanghai, 200030, People’s Republic of China }

\author{Zhu Liu}
\affiliation{Max Planck Institute for Extraterrestrial Physics, Giessenbachstrasse 1, D-85748, Garching, Germany}

\author{Youli Tuo}
\affiliation{Institut für Astronomie und Astrophysik, Universität Tübingen, Sand 1, D-72076 Tübingen, Germany}

\author{Yuanle Yao}
\affiliation{School of Physics and Technology, Wuhan University, Wuhan 430072, People’s Republic of China}

\author{Xinwu Cao}
\affiliation{Institute for Astronomy, School of Physics, Zhejiang University, 866 Yuhangtang Rd, Hangzhou, 310058, People’s Republic of China}

\begin{abstract}
The power-law emission and reflection component provide valuable insights into the accretion process around a black hole. In this work, thanks to the broadband spectra coverage of \emph{the Nuclear Spectroscopic Telescope Array}, we study the spectral properties for a sample of low-mass black hole X-ray binaries (BHXRBs). We find that there is a positive correlation between the photon index $\Gamma$ and the reflection fraction $R$ (the ratio of the coronal intensity that illuminates the disk to the coronal intensity that reaches the observer), consistent with previous studies, but except for MAXI J1820+070. 
It is quite interesting that this source also deviates from the well-known ``V"-shaped correlation between the photon index $\Gamma$ and the X-ray luminosity log$L_{\rm X}$, when it is in the bright hard state. More specifically, the $\Lambda$-shaped correlation between $\Gamma$ and log$L_{\rm X}$ is observed, as the luminosity decreases by a factor of 3 in a narrow range from $\sim 10^{38}$ to $10^{37.5}$ $\rm erg~s^{-1}$. Furthermore, we discover a strong positive correlation between $R$ and the X-ray luminosity for BHXRBs in the hard state, which puts a constraint on the disk-corona coupling and the evolution.
\end{abstract}

%

\section{Introduction}
\label{sec:intro}

The primary X-ray emission in X-ray binaries and active galactic nuclei (AGNs) produced via inverse Compton scattering of the soft photons in the hot compact region, often referred to as corona, is described by a power law with a high energy cutoff \citep{done2007,you2012,yuan2014}. A reflection component, which is believed to be produced by the illumination of the reflecting medium by the primary X-ray, is also observed. The reflection component is composed of fluorescent emission lines and a hump in the range of $\sim$20--40\,keV. We use the photon index $\Gamma$ to describe the slope of the primary X-ray, which indicates the fraction of the energy from the soft seed photons deposited into the corona. The 
reflection fraction $R$ is used to indicate the fraction of the primary X-ray intercepted by the reflection medium, namely, the relative cold disk \citep{gar2013}. The correlations between the X-ray properties (e.g., $\Gamma$ and broad-band luminosity $L_{\rm {X}}$) and $R$ imply the geometric and physical properties of the accretion flow around the central object \citep{wu2008,plant2014,you2021}. 

The positive correlation between $\Gamma$ and $R$ was found for the black hole X-ray binary (BHXRB) GX 339-4 in the hard state by fitting \emph{Ginga} spectra \citep{ued1994}. \citet{zdz1999} reported a similar correlation in Seyfert galaxies and in the hard state of four binary systems, and provided evidence for the reflective medium as the dominant source for seed photons to the corona. The correlation was also studied by analyzing \emph{RXTE}/PCA data for GX 339-4 and Cyg X-1 hard state \citep{gil1999, rev2001}, and for the sample of low-mass X-ray binaries (LMXBs; \citealt{ste2016}). However, the physical interpretation of this correlation is still under debate, which might be either the evolution of the inner radius \citep{zdz1999, qia2017, liu2022} in the scenario of the truncated disk, or the bulk velocity of the corona with respect to the disk \citep{bel1999}. 
As for the correlation between $L_{\rm {X}}$ and $R$, it was found that the $R$ gradually increases as the source luminosity rises for GX 339-4 in the hard state, which was interpreted in the scenario of the truncated disk \citep{plant2014}. However, a statistical study of the correlation between $L_{\rm {X}}$ and $R$ for a sample of LMXBs is required, in order to not only verify the observational correlation but also well understand the physical origin behind this correlation. 

The Nuclear Spectroscopic Telescope Array (\emph{NuSTAR}) is the first focusing telescope above 10\,keV \citep{har2013}. The high signal-to-noise ratio and broad energy band provide a good opportunity to better study the primary X-ray and the reflection hump, compared to previous spectral studies \citep{dia2020, dra2020, con2021,  marino2021,tripathi2021,feng2022,prabhakar2022,rout2022}. 
The correlation between $\Gamma$ and $R$ by analysing {\emph{NuSTAR}} spectra has been recently reported for Seyfert galaxies \citep{ezh2020, pan2020} and radio galaxies \citep{kan2020}.

The motivation of this work is to statistically study the correlations between the X-ray properties ($\Gamma$ and $L_{\rm {X}}$) and $R$, in a sample of stellar-mass black hole systems in the hard state, using \emph{NuSTAR} archived data. This paper is  organized as follows. In section \ref{sec:obs}, we present the sample selection and data reduction. In section \ref{sec:spe}, we give the spectral analysis and the best-fit results. Finally, the discussions and conclusions are presented in section \ref{sec:dis}. 


\section{Sample and Data Reduction} \label{sec:obs}

In this work, we adopt the \emph{NuSTAR} sample of 17 LMXBs  which is constructed by \citet{yan2020}, consisting of 165 observations downloaded from \emph{HEASARC}. The data reduction was performed through the \uppercase{nupipeline} task of \uppercase{nustardas} included in HEAsoft 6.29. 

We set \texttt{saamode=optimized} and \texttt{tentacle=yes} to filter passages through the South Atlantic Anomaly, and set \texttt{arfmlicorr=yes} to correct the temperature dependence of the multilayer insulation (MLI) for the FPMA data \citep{madsen2020}. For the bright sources, we use \texttt{STATUS==b0000xxx00xxxx000} to reprocess the data \footnote{https://heasarc.gsfc.nasa.gov/docs/nustar/analysis/}.
\uppercase{caldb} version 2021115 is used for the calibration. We extract the source spectra within a uniform radius of 90$\arcsec$ at the source position, and the background spectra within an annulus defined between 180$''$ and 200$''$.

\begin{figure*}
   \centering
    \includegraphics{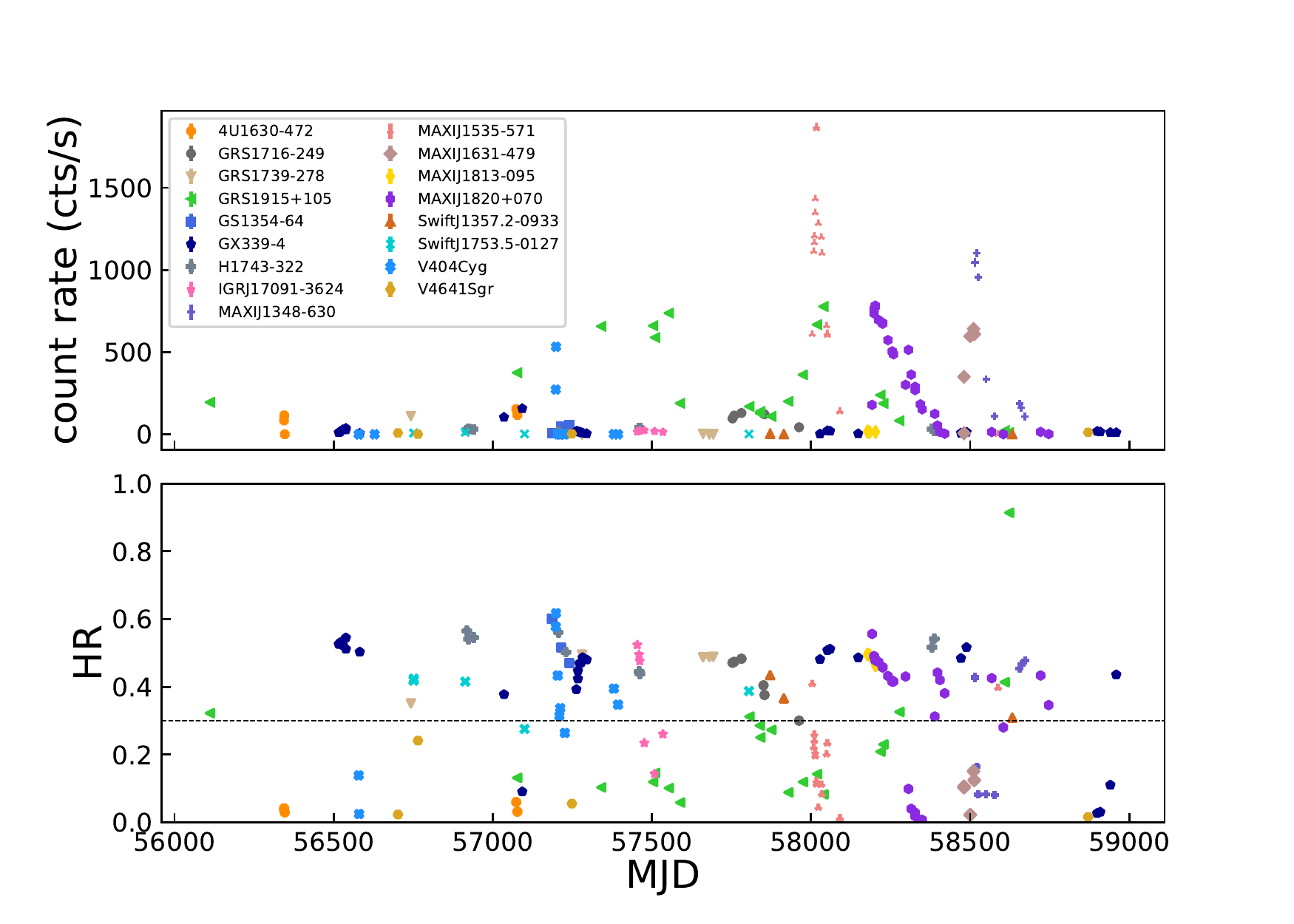}
   \caption{Overview of all observations, except the observation of 4U 1630-472 on MJD 56723 and the observation of H1743-322 on MJD 57251 due to background counts dominating their spectra. The net count rate within 3--78\,kev and HRs are shown in the upper and lower panels, respectively. The HR is defined as the count rate within the energy band of 10--78\,keV divided by the count rate within 3--10\,keV. Here, We only show an overview of \uppercase{fpma} for clarity.}
    \label{fig:lc_hness}%
    \end{figure*}

An overview of all data is presented in Figure \ref{fig:lc_hness}, in which the observation of 4U 1630-472 on MJD 56723 (30001016002) and the observation of H1743-322 on MJD 57251 (80002040006) are excluded, due to the spectra being dominated by the background. We plot the count rate within 3--78\,keV and the hardness ratio (HR) defined as the 10--78\,keV count rate divided by the 3--10\,keV count rate in the upper and lower panels, respectively. In this work, we aim to study the correlations between the X-ray primary and the disk reflection of LMXBs in the hard state. We set HR $>$ 0.3 as the selection criterion for the hard state. In this condition, observations of 4U 1630-472, MAXI J1631-479, and V4641 Sgr are excluded.
Furthermore, we also exclude observations for which the spectral fits do not require the reflection component, in the sense that those spectra can be well described by an absorbed power-law model, or an absorbed power-law plus {\tt {diskbb}} model.

Additionally, the observations of V404 Cyg on MJD 57198 (ObsID: 90102007002) and MJD 57199 (ObsID: 90102007003), and GRS 1915+105 on MJD 58608 (ObsID: 90501321002) and MJD 58623 (ObsID: 30502008002), are excluded, the spectra of which are quite complex and cannot be well fitted in our preliminary analysis with the reflection model ($\chi^2_{\nu}\gg2$). For V404 Cyg in the 2015 outburst, strong  flares were observed, which were thought to be related to transient jet activity of ejecting plasma close to a BH. The corresponding spectral fitting to the \emph{NuSTAR} observations on MJD 57198 and 57199 suggested that an X-ray source is close to a BH with a height smaller than 6 $R_{\rm g}$ and its spectra are extremely hard with $\Gamma \sim 1.4$ \citep{walton2017}. For GRS 1915+105, the \emph{NuSTAR} observations indicated that it was also in the flaring state on MJD 58608 and 58623, given the clear detection of the absorption features in the spectral residual, which were attributed to the accretion disk winds \citep{2020kol,ara2020}. 
Therefore, the final sample in this work is composed of 12 LMXBs with 69 observations in total. Detailed information is provided in Table \ref{tab:col} and Table \ref{tab:obs_detail}. 

\section{Spectral analysis} \label{sec:spe}

We perform the spectral fitting in \uppercase{xspec} 12.11.1 \citep{arn1996}. We account for the Galactic absorption using {\tt TBabs} \citep{wilms2000}. 
The column density ($N_\mathrm{H}$) was fixed to the values presented in Table \ref{tab:col}, most of which are obtained from the previous $NuSTAR$ spectral analysis in the literature. We adopt a relativistic reflection model {\tt relxillCp} \citep{gar2013, dau2014, gar2014}, to account for the reprocessed emission between the disk and corona. This reflection model contains both the direct Comptonization from the corona and its reflection on the disk. Therefore, {\tt {TBabs*relxillCp}} in XSPEC notation, is used to fit spectra in our sample. A possible disk component ({\tt diskbb}) and a nonrelativistic reflection model {\tt xillverCp} will be added if they can significantly improve the fitting.
The nonrelativistic reflection component {\tt xillverCp} may originate from the distant disk reflection. So, the ionization parameter is fixed at a low value with $\rm log \xi = 0.5$, for simplicity. And the Comptonization parameters (the electron temperature and the photon index) in {\tt xillverCp} are linked to the ones in {\tt relxillCp} \citep{buisson2019,you2021}. Note that, for the reflection model {\tt relxillCp}, no specific geometry of the Comptonization source is assumed, but with an artificial broken power-law emissivity. Alternative reflection models with significant findings of spectral complexity are also available in the literature \citep{mahmoud2018,aaz2021a,aaz2021b,kawamura2022}.

\begin{table*}
\caption{Basic information on the BHXRBs} \label{tab:col}
\centering
\begin{tabular}{lccccl} 
\hline
\noalign{\smallskip}
Source &$N_\mathrm{H}$ ($10^{22}$ $cm^{-2}$) &
Inclination ($^{\circ}$) &Mass $(M_{\odot})$&
Distance (kpc)&Ref.\\
\noalign{\smallskip}
\hline
\noalign{\smallskip}
GRS 1716-249       & 0.6  &-              & -  & 6&[1]  -   -  [2] \\
GRS 1739-278       & 2.3    &-              & -  & 7.5&[3]  -   -  [4] \\
GRS 1915+105       & 6.11   &$60\pm5$       &12.4 & 8.6&[5] [6] [6] [6] \\
GS 1354-64         & 0.7    &$\le79$        & 7.6& 7 &[7] [8] [8] [9] \\
GX 339-4           & 0.41  &37-78          & 5.9& 9  &[10][11][11][11] \\
H1743-322         & 2.3  &-              & -  & 8.5&[12][13] -  [13] \\
IGR J17091-3624    & 1.59  &-              & -  & 12  &[14] -   -   [27] \\
MAXI J1348-630     & 0.86 &-              & -  & 2.2&[15] -   -  [16] \\
MAXI J1535-571     & 5.5  &-              & -  & 4.1&[17] -   -  [18] \\
MAXI J1813-095     & 0.82  &-              & -  & 6  &[19] -   -   [26] \\
MAXI J1820+070     & 0.15 &$63\pm3$       & 8.48&2.96&[20][21][28][21]\\
Swift J1753.5-0127 & 0.2 &$\ge40$        & 7.4&8.42&[22][23][24][25]\\
\noalign{\smallskip}
\hline
\end{tabular}\\

[1]\citet{bha2019};
[2]\citet{saikia2022}; [3]\citet{mil2015}; [4]\citet{yan2017}; [5]\citet{mil2013}; [6]\citet{rei2014}; [7]\citet{ei2016}; [8]\citet{cas2009}; [9]\citet{gan2019}; [10]\citet{wj2018}; [11]\citet{hei2017}; [12]\citet{chand2020}; [13]\citet{ste2012}; [14]\citet{xu2017}; [15]\citet{sudip2021}; [16]\citet{cha2021}; [17]\citet{dong2022}; [18]\citet{cha2019}; [19]\citet{jiang2022}; [20]\citet{utt2018}; [21]\citet{atr2020};
[22]\citet{shaw2016}; [23]\citet{neu2014}; [24]\citet{sha2016}; [25]\citet{gan2019}; [26]\citet{jan2021}; [27]\citet{iye2015}; [28]\citet{torres2020}
 
\end{table*}

As for {\tt relxillCp}, the inclination angle $i$ is left free in the spectral fitting, except GRS 1915+105, MAXI J1820+070, and GX 339-4. The inclination angles of the first two sources are adopted from the measurements of jet inclination angles \citep{rei2014,atr2020}, i.e., $i = 60^\circ$ for GRS 1915+105 and $i = 63^\circ$ for MAXI J1820+070. However, we notice that there is a large misalignment ($>40^\circ$) between the inclination angles of the jet and binary orbit in MAXI J1820+070 \citep{pou2022}. As for GX 339-4, the binary inclination is constrained to be $37^\circ < i < 78^\circ$ from the VLT/X-shooter observations. Moreover, as suggested by \cite{aaz2019}, the inclination of this source is unlikely to be high, which is supported by the absence of the disk winds \citep{ponti2012} and the track on the  hardness–intensity diagram \citep{munoz2013}. In the preliminary spectral fits of GX 339-4, the inclination is pegged at a lower limit. Therefore, the inclination angle of this source is fixed at $i=40^\circ$ in the following fits. In this work, we aim at determining the connection between the corona and accretion disk, while the robust estimate of the BH spin $a_*$ for an individual source of the sample is beyond the scope of this paper. Therefore, for simplicity, we fix the BH spin $a_* = 0.998$ for the \emph{NuSTAR} sample here.
The photon index $\Gamma$, reflection parameter $R$, the electron temperature $kT_{\rm e}$, the iron abundance $A_{\rm{Fe}}$ and normalization were free parameters in the spectral fits. 


For a given source, the two free parameters, i.e., the inclination of the binary system and the Fe abundance of the accretion disk, are expected to be constant. Therefore, we perform joint fits to multi-epoch spectra for a given source, in the sense that the two parameters between the multi-spectra are linked. Given that (i) the complexity in configuring the xcm files for the joint fits with the complex reflection model, and (2) the reported rip occurring in early 2016 in the MLI associated with \uppercase{fpma}, which resulted in an increased photon flux in \uppercase{fpma} spectra of bright sources \footnote{https://heasarc.gsfc.nasa.gov/docs/nustar/nustar\_faq.html\#MLI}, we will only use the \uppercase{fpmb} spectra in the energy band of 3--78\,keV, for each observation. For MAXI J1820+070, due to the complex fitting model (the disk component, the relativistic reflection and the nonrelativistic reflection) and there being 18 spectra in total, we divide these spectra into two data sets: one for the rising hard state before MJD=58303, and the other one for the decaying hard state after MJD = 58303 \citep{buisson2019}. The iron abundance of the second data set is fixed at the best-fitting value of the second data sets, in order to keep this parameter constant during the outburst in 2018. 

The best-fitting results are obtained by implementing a Markov Chain Monte Carlo (MCMC) algorithm in XSPEC to create a chain of parameter values whose density gives the probability distribution for that parameter. The analysis of the probability distributions derived from the MCMC chains is implemented using the corner package \cite{Foreman2013}. The best-fit results for each free parameter are taken as the medians of the corresponding posterior distributions, and the error bars correspond to the 68\% confidence interval ($1\sigma$ uncertainty).

The best-fitting results are presented in Table 3. Once the best-fitting result is derived, we then use the convolution model \emph{cflux} to estimate the unobserved 3--78\,keV X-ray fluxes of the {\tt relxillCp} component. 



\begin{figure*}
\includegraphics[width=\linewidth]{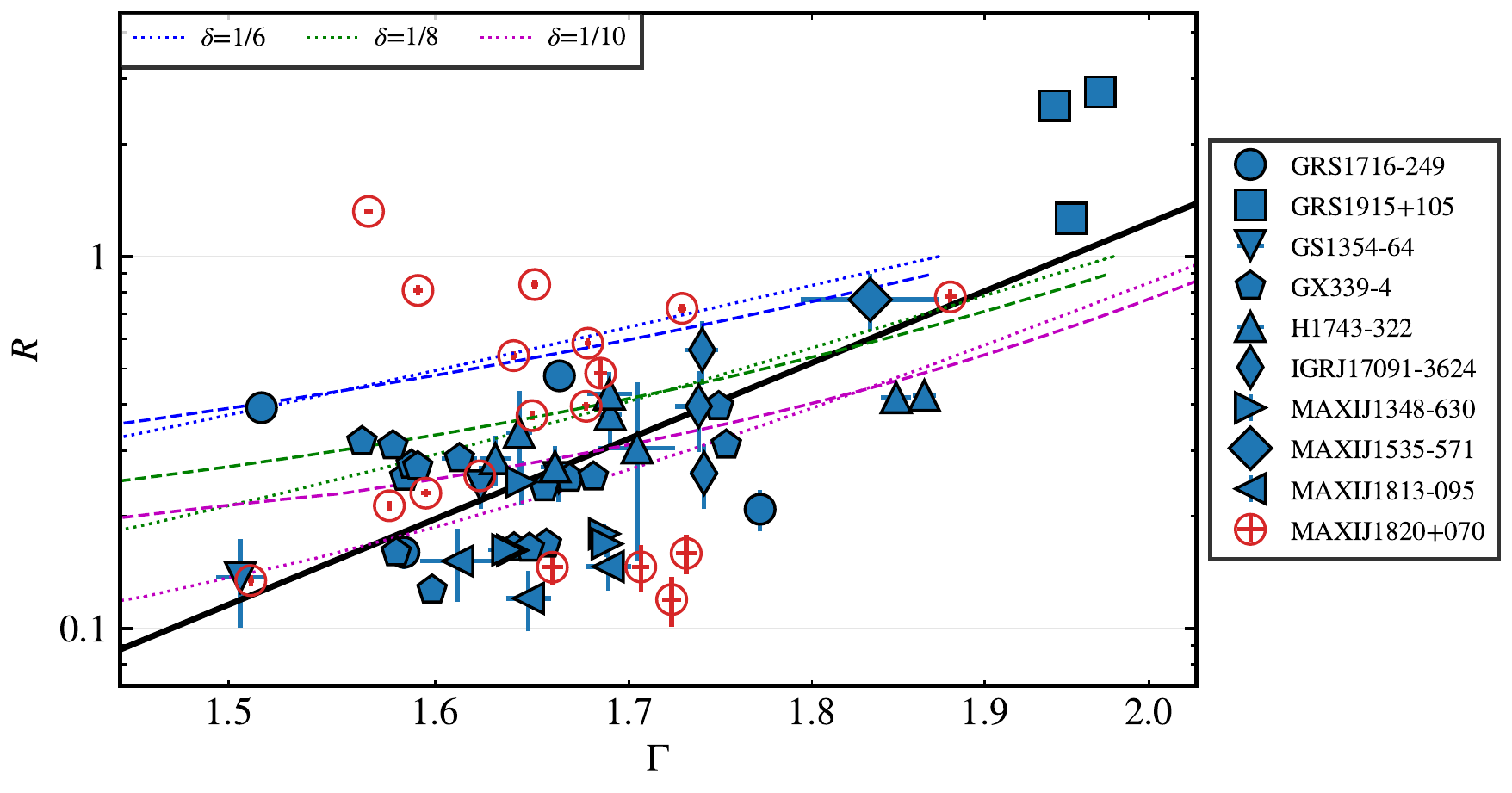}
\caption{The relationship between $\Gamma$ and $R$. GRS 1739-278 and Swift J1753.5-0127 are not included, due to that the spectra can be well fitted by a simple Comptonization model without the presence of the reflection features. The solid line denotes the best-fitting result excluding MAXI J1820+070 (open red circle). We also calculate the relationship between $\Gamma$ and $R$ based on the models proposed by \citet[shown in dotted lines]{zdz1999} and \citet[shown in dashed lines]{bel1999}. The blue, green, and purple colors correspond to $\delta=1/6$, $\delta=1/8$, and $\delta=1/10$, respectively, in Eq. (2).
}
\label{fig:r_gamma}
\end{figure*}

\begin{figure*}
\includegraphics[width=\linewidth]{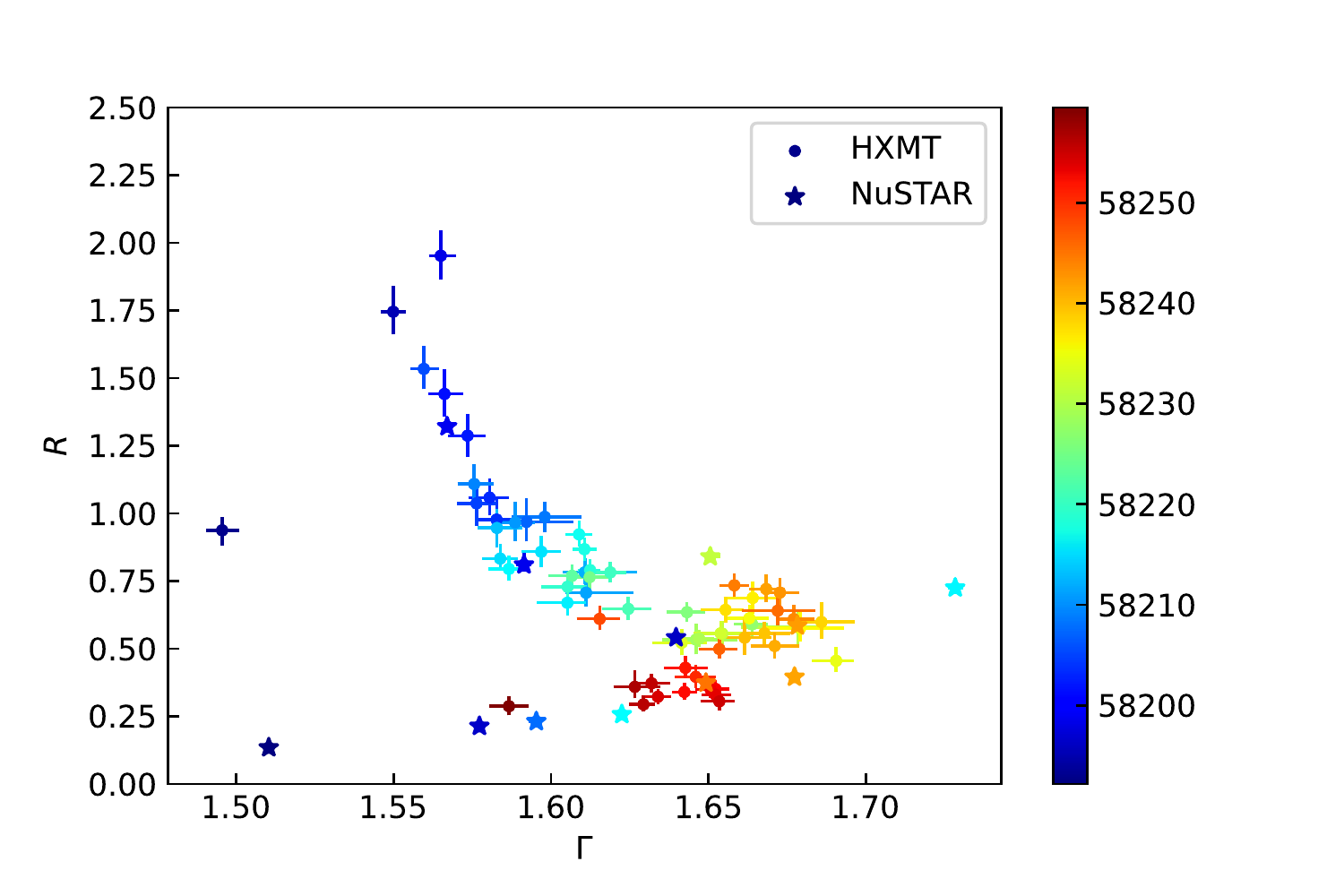}
\caption{The relationship between $\Gamma$ and $R$ for MAXI J1820+070 in the rising hard state during the outburst rise in 2018 (before MJD = 58303), in which the estimations from the spectral fitting to both \emph{NuSTAR} (blue points) and \emph{Insight}-HXMT (red points) observations are included. The color of each point
corresponds to the observation MJD, with the color bar on the right.}
\label{fig:1820_ref_gamma}
\end{figure*}

\begin{figure*}
\includegraphics[width=\linewidth]{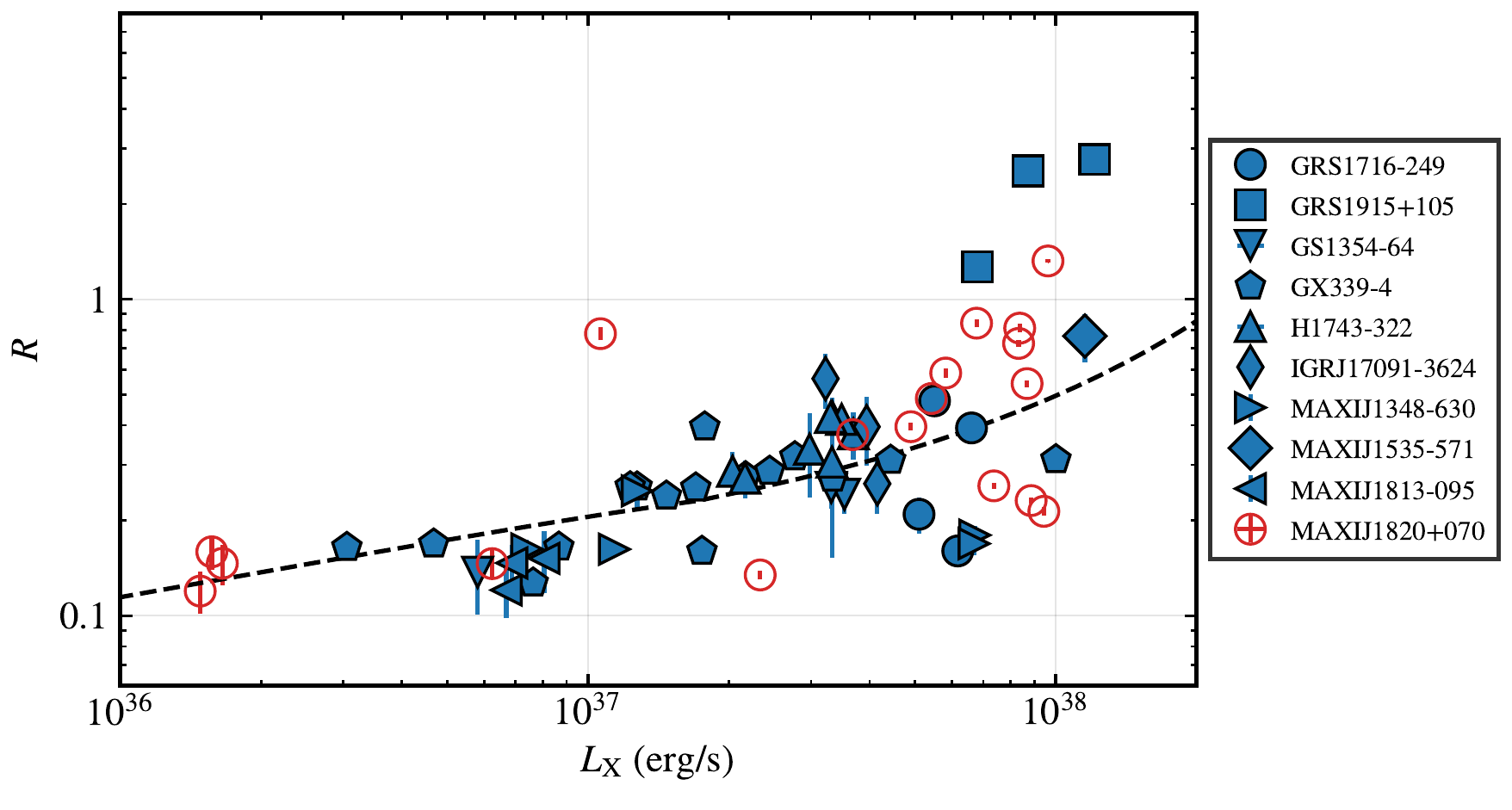}
\caption{Correlation between $R$ and the X-ray luminosity $L_{\rm X}$ of the {\tt relxillCp} component in 3--78\,keV for the \emph{NuSTAR} sample. GRS 1739-278 and Swift J1753.5-0127 are not included, due to that the spectra can be well fitted by a simple Comptonization model without the presence of the reflection features.}
\label{L_R}
\end{figure*}

\begin{figure*}
\includegraphics[width=\linewidth]{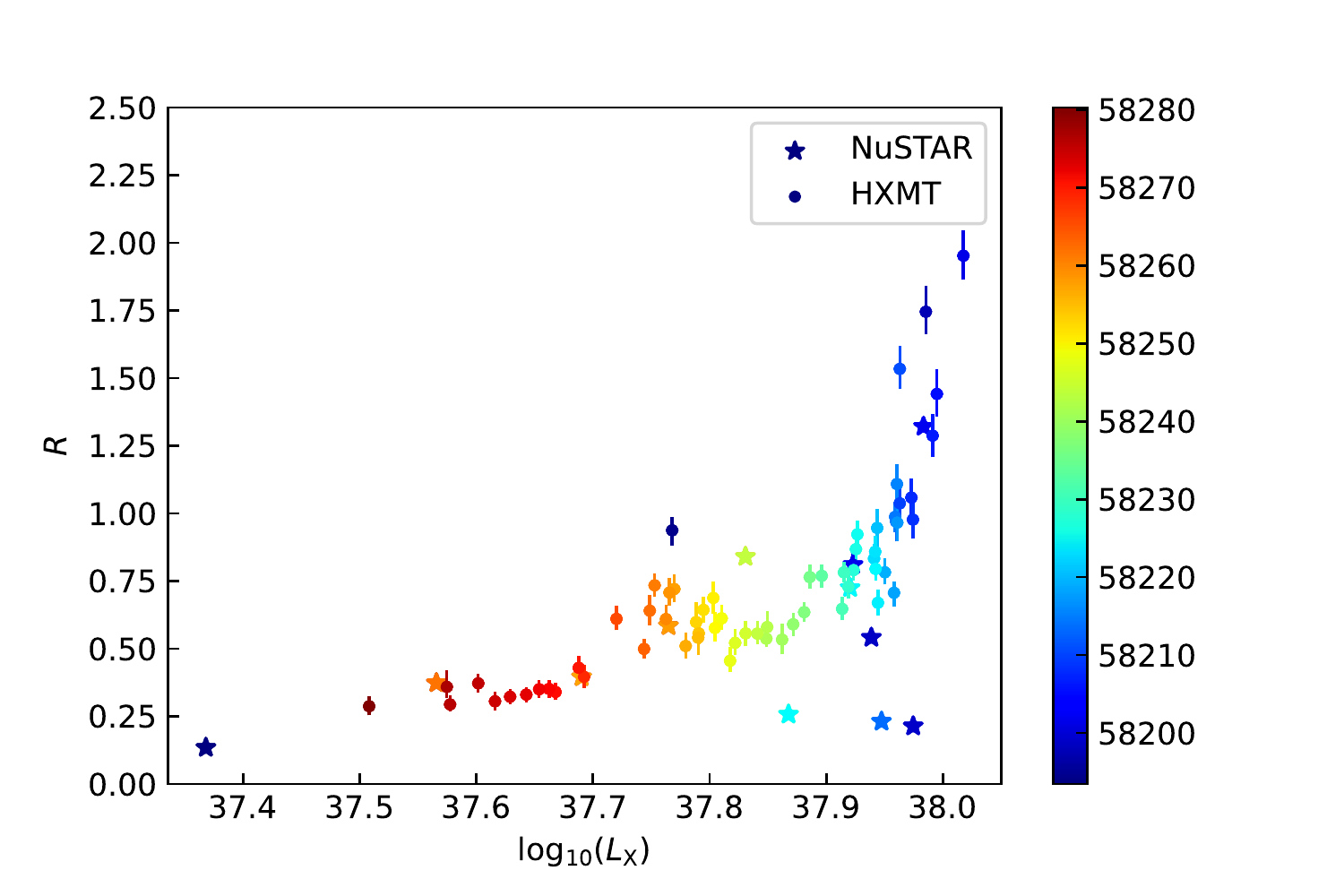}
\caption{Correlation between $R$ and the X-ray luminosity $L_{\rm X}$ of the {\tt relxillCp} component in 3--78\,keV for MAXI J1820+070 in the rising hard state during the outburst rise in 2018 (before MJD = 58303), in which the estimates from the spectral fits to both \emph{NuSTAR} (star points) and \emph{Insight}-HXMT (circle points) observations are included. The color of each point corresponds to the observation MJD, with the color bar on the right. }
\label{fig:1820_ref_lx}
\end{figure*}

\begin{figure*}
\includegraphics[width=\linewidth]{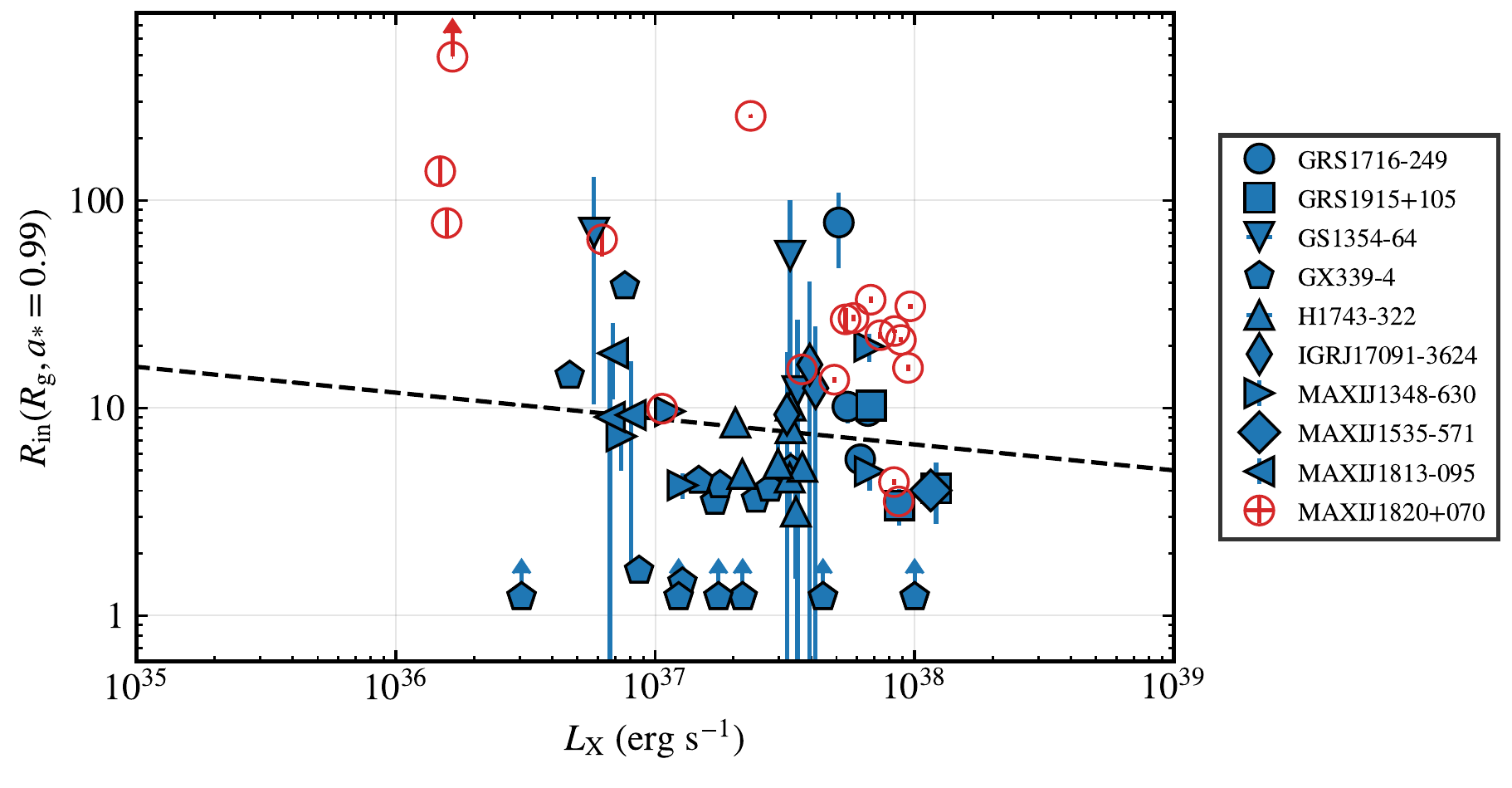}
\caption{Correlation between $R_{\rm in}$ and the X-ray luminosity $L_{\rm X}$ of the {\tt relxillCp} component in 3--78\,keV for the \emph{NuSTAR} sample. GRS 1739-278 and Swift J1753.5-0127 are not included, due to that the spectra can be well fitted by a simple Comptonization model without the presence of the reflection features.}
\label{L_Rin}
\end{figure*}

\begin{figure*}
\includegraphics[width=\linewidth]{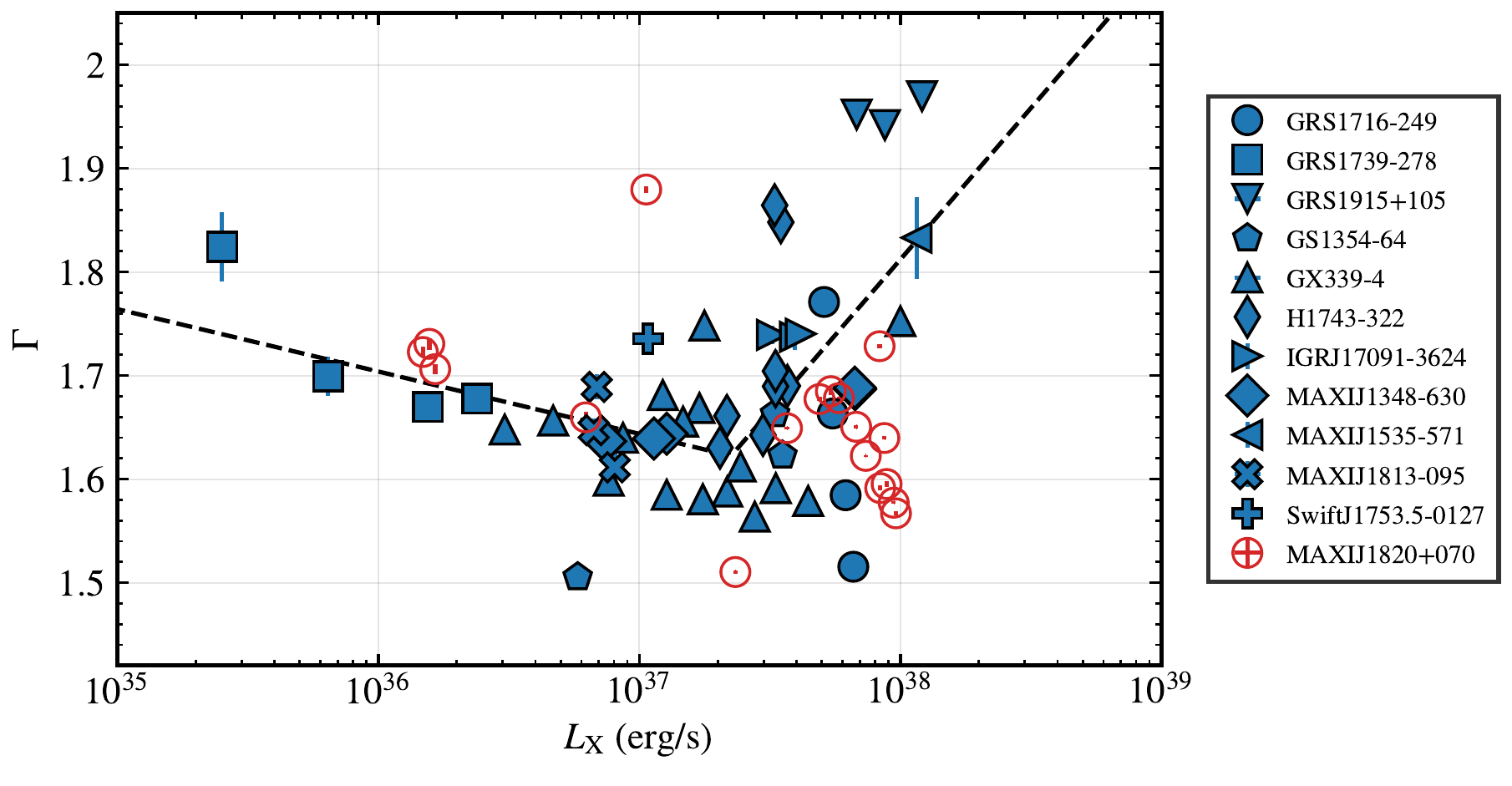}
\caption{Correlation between photon index $\Gamma$ and the X-ray luminosity $L_{\rm X}$ of the {\tt relxillCp} component in 3--78\,keV for the \emph{NuSTAR} sample. The different colors and symbols indicate the different sources, just the same as the figure legend in Figure \ref{fig:lc_hness}}
\label{L_gamma}
\end{figure*}

\begin{figure*}
\includegraphics[width=\linewidth]{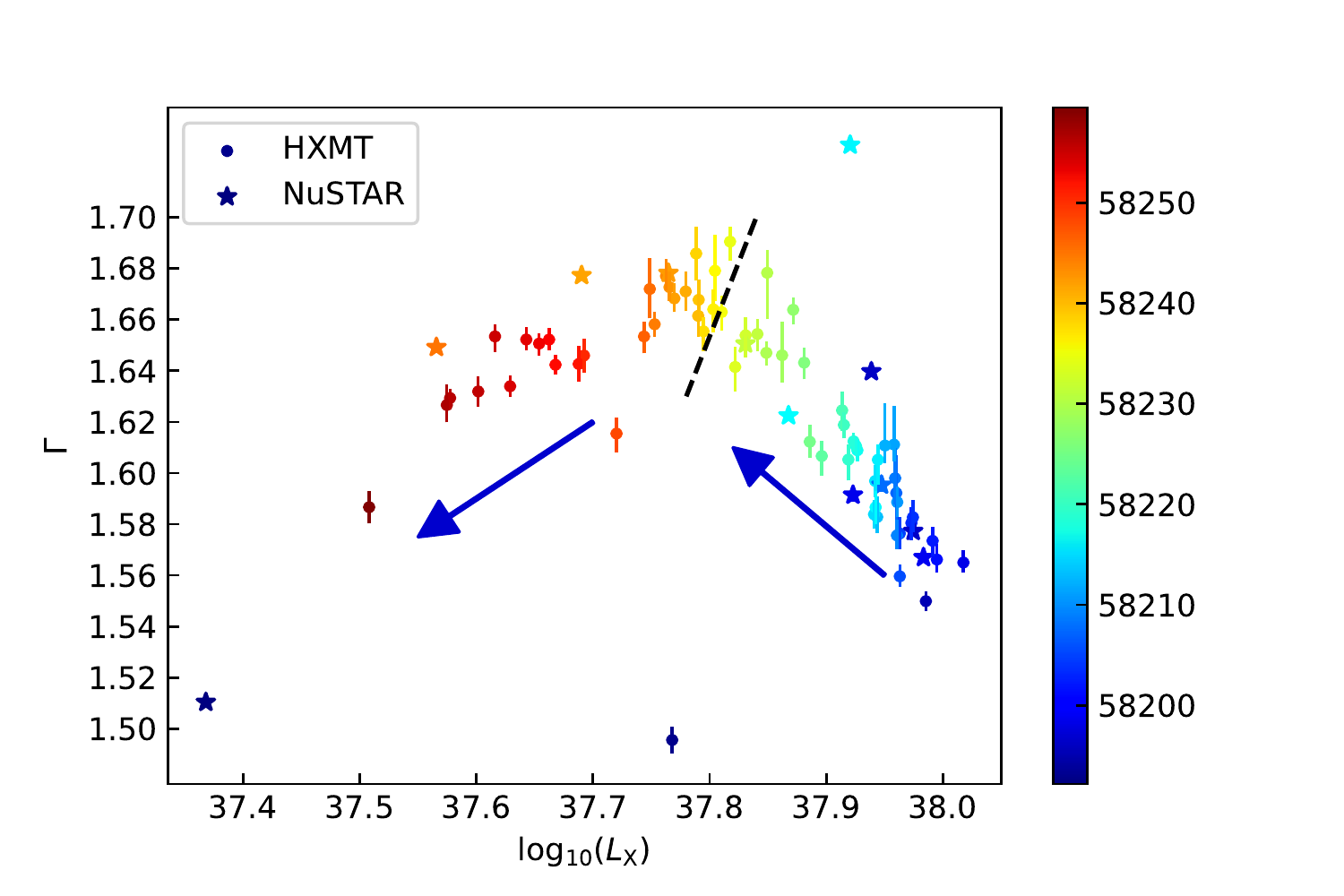}
\caption{The relationship between $\Gamma$ and the X-ray luminosity $L_{\rm X}$ of the {\tt relxillCp} component in 3--78\,keV for MAXI J1820+070 in the rising hard state during the outburst rise in 2018 (before MJD = 58303), in which the estimates from the spectral fits to both \emph{NuSTAR} and \emph{Insight}-HXMT observations are included. The color of each point corresponds to the observation MJD, with the color bar on the right. The dashed line roughly corresponds to MJD = 58250.}
\label{fig:1820_gamma_lx}
\end{figure*}

\begin{figure*}
\includegraphics[width=\linewidth]{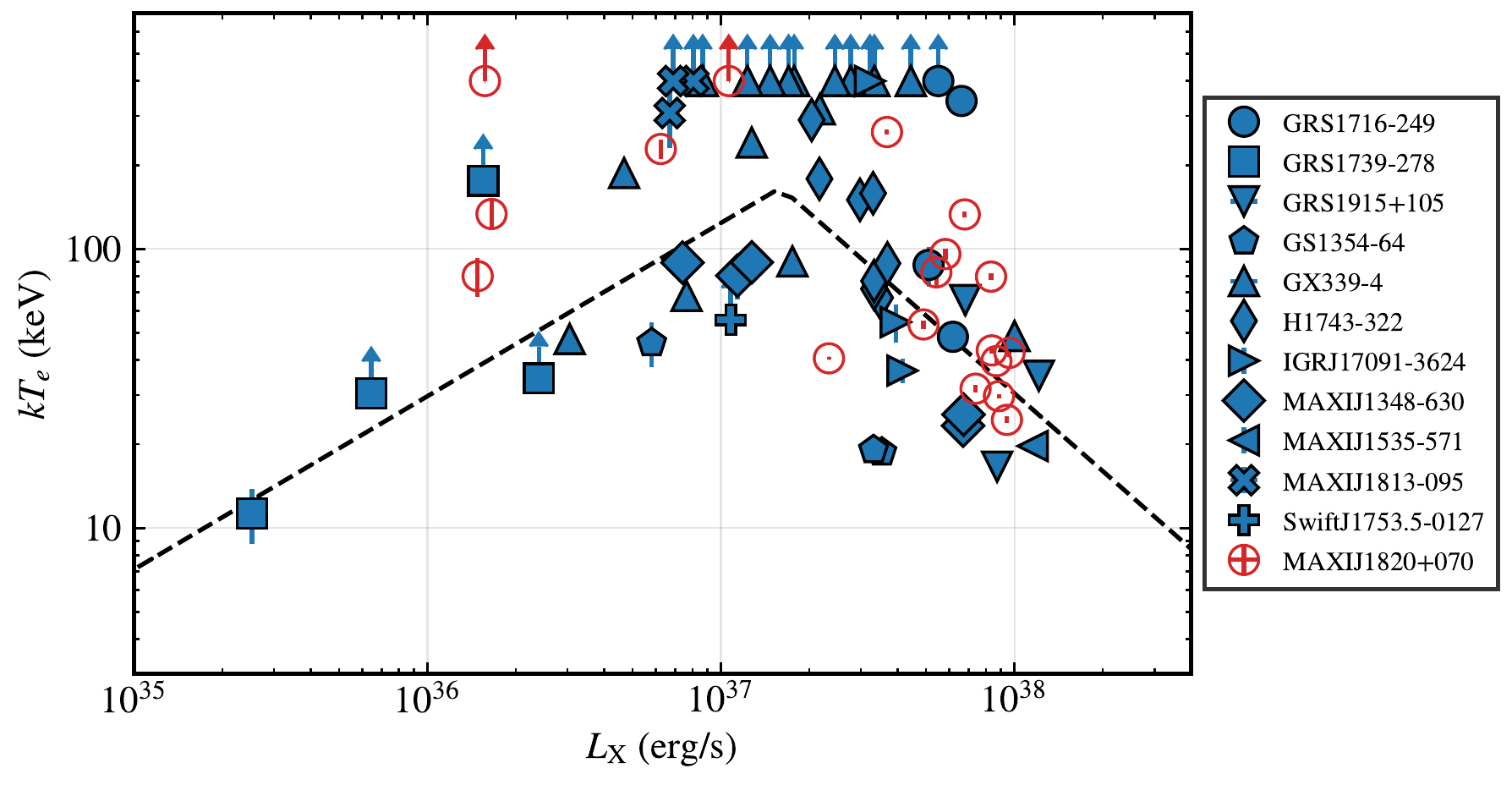}
\caption{Correlation between the electron temperature $kT_{\rm e}$ and the X-ray luminosity $L_{\rm X}$ in 3--78\,keV for the \emph{NuSTAR} sample.}
\label{L_kte}
\end{figure*}


\section{Results and Discussion}
\label{sec:dis}
\subsection{Disk-Corona Relation}
 The hard X-ray from the corona illuminates the disk, which is reprocessed within the disk to produce the reflection spectrum \citep{george1991,gar2013,you2021,klepczarek2023}. Therefore, the reflection spectrum is a crucial ingredient to study the relation between the corona and disk \citep{you2021}. We then explore the disk-corona relation through the correlations between different spectral parameters.
\subsubsection{The $\Gamma$-$R$ relation}
\label{subsec:cor}

For GRS 1739-278 and Swift J1753.5-0127, the spectra can be well fitted by a simple Comptonization model {\tt nthcomp} \citep{aaz1996} without the presence of the reflection features. This means the reflection fraction cannot be well constrained by the data. Therefore, excluding these two sources, we examine the correlation between $R$ and $\Gamma$ using Spearman rank-order test. In order to take into account uncertainties of the parameters, we perform Monte Carlo sampling of the $R$ and $\Gamma$ \citep{curran2014}. We notice that some observations of MAXI J1820+070 apparently deviate from this correlation with a large scatter. 
The complexity of this source will be discussed below. If MAXI J1820+070 is excluded from the sample, we obtain a correlation between $R$ and $\Gamma$, with the Spearman's coefficient of $0.53$ at a significance of 3.78 $\sigma$. 


Furthermore, in order to better understand the physical origin of the correlation, we need to derive the form of the correlation. \cite{zdz1999} studied the correlation between $\Gamma$ and the $R$ by fitting a few phenomenological functions to the data. And, they found that the best model is a power law, $R = u \Gamma ^\nu$ with $u = (1.4 \pm 1.2) \times 10^{-4}$, $\nu = 12.4 \pm 1.2$. Here, we use the \texttt{linmix} algorithm \citep{kelly2007} implemented in Python \footnote{\url{https://github.com/jmeyers314/linmix}} to fit the above correlation with a linear function $\log R = \nu \times \log \Gamma+ u$. The best-fitting results are  $\nu=8.20\pm1.20$, $u=-2.38\pm0.27$   (see also Figure \ref{fig:r_gamma}).



The reflection fraction could be approximated as the ratio of the luminosity of the primary X-ray source with unit luminosity incident on the disk to the one radiated outward to infinity, 
\begin{equation}
    R = L_{\rm d}/(1-L_{\rm d}),
\end{equation}
where $L_{\rm d}$ is the incident luminosity over the disk \citep{zdz1999}. And the photon index $\Gamma$ can be estimated from the Compton amplification factor of the thermal Comptonization $A$ \citep{bel1999},
\begin{equation}
    \Gamma \approx 2.33(A-1)^\delta.
\end{equation}
The Compton amplification factor is defined as $A \equiv 1/L_{\rm s}$ , where $L_{\rm s}$ is the power in seed photons scattered in the sphere (assuming a unit optical depth).

In the scenario of the truncated disk, both the reflection fraction $R$ and the Compton amplification factor $A$ depend on the disk truncation radius $d$, which can be estimated from Eq. (1) and (2) of \cite{zdz1999}.
In the scenario of the outflowing X-ray source, assuming the geometry is steady, both $R$ and $A$ also depend on the outflowing velocity of the corona, which can be determined from Eq. (3) and (7) of \cite{bel1999}.
We plot the predicted correlations between $R$ and $\Gamma$ for these two scenarios, in the case of $\delta=1/6$, $\delta=1/8$, and $\delta=1/10$ (see the dotted and dashed lines). It can be seen that both scenarios predict positive correlations between $R$ and $\Gamma$, which are qualitatively consistent with the observed correlation (except for MAXI J1820+070) in the \emph{NuSTAR} sample. 

Some observations of MAXI J1820+070 apparently deviate from the positive $R$-$\Gamma$ correlation above. It seems that there is a negative correlation between $R$ and $\Gamma$ in MAXI J1820+070. 
 We notice that \emph{Insight}-HXMT also monitored this source with higher cadence and in a broader energy band (2--150\,keV) during the outburst in 2018, which was comprehensively studied in \cite{you2021}. In order to study the evolution of MAXI J1820+070, we also plot the $\Gamma$ and $R$ obtained from \cite{you2021} in Figure \ref{fig:1820_ref_gamma}. \footnote{At the time of preparing the publication of \cite{you2021}, {\tt relxill} v\ 1.4.0 was the latest available version, with the bug fixed in v\ 1.4.3, see \url{http://www.sternwarte.uni-erlangen.de/~dauser/research/relxill/}. In this work, using the latest version of relxill, the spectral fits to \emph{Insight}-HXMT spectra are updated.} Combining the results from both \emph{NuSTAR} and \emph{Insight}-HXMT, it is evident that (i) the reflection fraction is indeed large and can reach up to $\sim 2$ during the peak of the rising hard state (around MJD 58200); (ii) there is indeed a negative correlation, where the reflection fraction rapidly decreases from $\sim$ 2.0 to 0.25, as $\Gamma$ increases in a narrow range from 1.55 to 1.7. The Spearman coefficient is -0.59 at a significance of 5.88 $\sigma$.

\subsubsection{The $R$-$L_{\rm X}$ correlation}
\label{subsec:r_lx}
We then study the possible correlation between $R$ and log$L_{\rm X}$ for this \emph{NuSTAR} sample. 
The Spearman coefficient of the correlation between $R$ and $L_\mathrm{X}$ is 0.66 at a significance of 5.02 $\sigma$. This positive correlation can be explained in the truncated disk model. It is hard to directly compute the Comptonization luminosity $L_{\rm C}$, but it is proportional to the luminosity $L_{\rm S}$ of the seed photons which are scattered in the X-ray source, i.e., $L_{\rm C} \propto L_{\rm S}$ \citep{pou2018}. The reflection fraction $R$ is computed with Equation (1), and $L_{\rm S}$ is computed using Equations (1) and (2) of \cite{zdz1999}. $L_{\rm S}$ is a function of the truncation radius $d$ of the accretion disk. The model prediction of the $R$--$L_\mathrm{X}$ correlation is plotted in \autoref{L_R}, as a dashed line. The observations roughly follow the model prediction below $\sim 5\times10^{37}~\rm erg~ s^{-1}$, and apparently deviate from it at higher luminosity.

The positive correlation between $R$ and $L_\mathrm{X}$ for MAXI J1820+070 is much more evident than the one for the sample study above. For clarity, we plot the estimates $R$ and log $L_{\rm X}$ in Fig. \ref{fig:1820_ref_lx}, including the results of both \emph{NuSTAR} and \emph{Insight}-HXMT during the rising hard state of the outburst in 2018. It can be seen that these combined estimates do show a positive correlation, with the reflection fraction rapidly increasing over one order of magnitude as the luminosity increases by a factor of 3. The Spearman coefficient is 0.8 at a significance of 9.4 $\sigma$.



In the truncated disk model \citep{esin1997,done2007,qiao2012}, it is proposed that the outer disk truncation radius decreases with the mass accretion rate, which leads to the softening of the spectrum. If assuming the inner Comptonization source (i.e., advection-dominated accretion flow) is static, the reflection fraction will increase with the mass accretion rate (i.e., the luminosity), given that the reflection fraction increases as the truncation radius decreases \citep{zdz1999}.
\cite{plant2014} carried out a systematic analysis of the reflection spectrum throughout three outbursts of GX 339-4. It was found that $R$ gradually increases as the source luminosity for this source in the hard state, which was interpreted in the scenario of the truncated disk. In \cite{demarco2017}, the X-ray reverberation lags in GX 339-4 were measured in the hard state, which were found to decrease as a function of X-ray luminosity.
In the truncated disk model, the decrease in the lag amplitude suggests a decrease in the relative distance between the disk and corona, which may lead to an increase in the reflection fraction. Therefore, the positive correlation $L_{\rm X}$-$R$ derived in this work, is consistent with previous spectral and timing studies.
Recently, in \cite{sridhar2020}, the \emph{RXTE} observations of of GX 339-4 in the 2002 and 2004 outbursts were reanalysed. They found that the reflection strengths in the 2002–2003 outburst with high X-ray flux is larger than the ones in the 2004–2005 outburst with low X-ray flux (see their Tables 2 and 3).

\subsubsection{The $R_{\rm in}$-$L_{\rm X}$ correlation}

In this work, the disk inner radius is estimated from the spectral fits with the reflection model. In Figure \ref{L_Rin}, we plot the evolution of the disk inner radius with the X-ray luminosity of the reflection model in 3--78\,keV, which shows that the disk is truncated in the hard state.
We fit the correlation with the linear function Eq. (\ref{bpl}), by replacing the $X$ and $Y$ with log$L_{\rm x}$ and log$R_{\rm in}$ and excluding all the lower limits of $R_{\rm in}$. The best-fit result is plotted as the solid line. The best-fitting slope is -0.12, which indicates a negative correlation. 
The Spearman coefficient of the correlation for the data is -0.04 at the significance of 0.25 $\sigma$. Due to the large scatter, we cannot draw a conclusion that a negative correlation between R and L exists among the sample.


By recalibrating the \emph{RXTE} data to a better precision, \cite{garcia2015} reanalyzed six composite RXTE spectra of GX 339–4 in the hard state, with the luminosity spanning by one order of magnitude. 
The reflection model {\tt relxill} was used to measure the inner disk radius. They found that the disk is slightly truncated by a few gravitational radii $R_{\rm g}$ at a few percent of Eddington, and the disk inner radius tends to decrease with the source luminosity. The same conclusion of the evolution of the disk inner radii with the source luminosities for GX 339-4 was also drawn in \cite{plant2015}. Recently, the same timing analysis was also applied to MAXI J1820+070 by \cite{demarco2021}. It was found that the frequency of thermal reverberation lags steadily increases during the outburst in 2018, implying that the relative distance between the X-ray source and the disk decreases (i.e., the decrease of the inner radius of the disk) as the source softens.

In this work, the disk component is required in the spectral fits to the NuSTAR data for some of the sources in the sample, e.g., MAXI J1820+070. The inner radius can also be estimated from the normalization of the {\tt diskbb} \citep{kubota1998}. However, in the spectral fits to \emph{Insight}-HXMT observations of MAXI J820+070, the estimated $R_{\rm in}$ from the diskbb components is larger than the ISCO, while the disk inner radius is fixed at ISCO \citep{you2021} in the relativistic reflection model, based on the spectral/timing results in \cite{kara2019}. Such a disagreement would indicate that the {\tt diskbb} components are only phenomenological descriptions of the complex spectra \citep{aazyb2022}. More importantly, the estimate of $R_{\rm in}$ from the {\tt diskbb} component also strongly depends on the boundary condition and the color hardening factor \citep{kubota2004,sridhar2020,aaz2022}. 

\subsection{Luminosity-dependent corona properties}
\subsubsection{$\Gamma$-$L_{\rm X}$ Correlation}
In this work, we do the spectral fitting for the \emph{NuSTAR} sample of LMXBs using the relativistic reflection model. The broadband luminosity and the corona properties, including the electron temperature and the photon index, can be well measured, so that we could statistically study the dependence of the corona properties on the X-ray luminosity.
For the accreting BH sources (including AGNs and BHXRBs), it was found that there is a ``V"-shaped correlation between the photon index $\Gamma$ and log$L_{\rm X}$ \citep{yamaoka2005,wu2008,yang2015,yan2020}, i.e.,
\begin{equation}
  Y(X) = \left\{
  \begin{array}{cc}
    f_1 (X-X_{c}) + Y_{c}  & \quad (X\leq X_{c}),  \\
    & \\
    f_2 (X-X_{c}) + Y_{c}  & \quad (X\geq X_{c}).
  \end{array} \right.
  \label{bpl}
\end{equation}
In this work, we investigate this correlation for the \emph{NuSTAR} sample of LMXBs. The derived $\Gamma$ and log$L_{\rm X}$ from the spectral fitting are plotted in Fig.\ref{L_gamma}. Apparently, there is a also ``V''-shape correlation between $\Gamma$ and log$L_{\rm X}$. Following \cite{yan2020}, we fit the correlation with the broken linear function Eq. (\ref{bpl}), by replacing the $X$ and $Y$ with log$L_{\rm x}$ and $\Gamma$. The best-fit result is plotted as the solid line in Fig.\ref{L_gamma}. The correlation is broken at the luminosity $\log L_{\rm X} = 37.35$, below which the slope is $-0.06$, and above which the slope is 0.29. The Spearman coefficients of the correlation for the data below and above this broken luminosity are -0.30 and 0.39 at the significances of 1.43 $\sigma$ and 1.91 $\sigma$, respectively. 
Due to the larger scatter between different sources, the negative/positive correlation below/above the broken luminosity are not very significant. This is consistent with previous studies on AGNs and BHXRBs \citep{wu2008,yang2015,yan2020}.

In Fig. \ref{fig:1820_gamma_lx}, we plot the estimates $\Gamma$ and log ($L_{\rm X}$) from both this work and \cite{you2021}, for MAXI J1820+070. Meanwhile, we associate each data point with the corresponding MJD, indicated as the color bar on the right. It can be seen that the \emph{NuSTAR} results are consistent with the \emph{Insight}-HXMT ones. More importantly, we notice an interesting ``$\Lambda$''-shaped correlation between the photon index $\Gamma$ and log$L_{\rm X}$, 
as the luminosity decreases by a factor of 3 in the narrow range from $\sim 10^{38}$ to $10^{37.5}$ $\rm erg~s^{-1}$. More specifically,  $\Gamma$ firstly increases as $L_{\rm X}$ decreases. But, after around MJD = 58250 (see the dashed line in Fig. \ref{fig:1820_gamma_lx}), $\Gamma$ starts to decrease as $L_{\rm X}$ continually decreases. We note that
\cite{wang2020} studied the evolution of the temporal properties of MAXI J1820+070 over the same time with \emph{Insight}-HXMT data sets. They also found the different behaviors of the HR, the fractional rms, and time lag before and after MJD 58,257, which suggests a transition occurred around this point.

\subsubsection{$kT_{\rm e}$-$L_{\rm X}$ Correlation}
It was also found that there is a ``$\Lambda$''-shaped correlation between the electron temperature $kT_{\rm e}$ and log$L_{\rm X}$ \citep{yan2020}. Note that \cite{yan2020} were not aimed at deriving the exact value of
$kT_{\rm e}$, but instead at the trend of the $kT_{\rm e}$-$L_{\rm X}$ correlation. So the spectral fitting was simply carried out by the use of the Comptonization model ({\tt nthcomp}) without considering the role of the reflection component. This is acceptable for the low-luminosity regime $<10^{37}$ $\rm erg~s^{-1}$, since the reflection emission in this case is systematically weak \citep{furst2016,beri2019}. However, the estimate of $kT_{\rm e}$ is probably affected by the relativistic reflection components, for the high-luminosity regime $>10^{37}$ $\rm erg~s^{-1}$, which is covered by the sample in this work \citep{fabian2015,garcia_dauser2015}. Therefore, it is essential for us to revisit the $kT_{\rm e}$-$L_{\rm X}$ correlation using the measurements from the spectral fitting with the relativistic reflection model. We plot the $kT_{\rm e}$ and $L_{\rm X}$ derived from the spectral fitting with the reflection model in Fig.\ref{L_kte}, which appear to still display the ``$\Lambda$''-shaped correlation between them. 
We fit the correlation with the broken linear function Eq. (\ref{bpl}), by replacing the $X$ and $Y$ with log$L_{\rm x}$ and $\log kT_{\rm e}$ and excluding all the lower limits of $kT_{\rm e}$. The best-fitting result is plotted as the dashed line. The correlation is broken at the luminosity log$L_{\rm X} = 37.20$, below which the slope is 0.62, and above which the correlation is $-0.93$. 
The Spearman coefficients of the correlation for the data below and above this broken luminosity are 0.50 and -0.61 at the significances of 1.42 $\sigma$ and 3.16 $\sigma$, respectively. This is qualitatively consistent with previous studies on BHXRBs \citep{yan2020}.


\section{Conclusions}
\label{sec:con}
In this work, we systematically study the evolution of the primary X-ray and the reflection component, for the \emph{NuSTAR} sample of LMXBs in the hard state. The photon index $\Gamma$ and the reflection fraction $R$ are well constrained. We find that $R$ is positively correlated with $\Gamma$ for the sample except for MAXI J1820+070. The positive $R-\Gamma$ correlation is in consistent with previous studies of LMXBs \citep{ued1994, zdz1999, ste2016} and AGNs \citep{dia2020, dra2020,con2021}. The $\Gamma$-$R$ correlation could be explained in the scenarios of both the truncated disk and the outflowing X-ray source. Then we calculate the X-ray luminosity of the reflection model between 3 and 78\,keV. It is interesting that there is a strong positive correlation between $R$ and the X-ray luminosity for BHXRBs in the hard state, which puts a constraint on the disk-corona coupling and the evolution. Moreover, We report ``V''-shaped correlation between $\Gamma$ and log$L_{\rm x}$ and ``$\Lambda$''-shaped correlation between $kT_{\rm e}$ and log$L_{\rm x}$, which is in agreement with previous results. Combining the \emph{NuSTAR} and \emph{Insight}-HXMT results which obtained from \citep{you2021}, we find a negative $R-\Gamma$ correlation and a ``$\Lambda$''-shaped correlation between $\Gamma$ and log$L_{\rm x}$ for MAXI J1820+070 in the bright hard state. The physical mechanism behind the particular behavior of MAXI J1820+070 is unknown. It challenges our knowledge of the accretion flow of black hole binary, and needs further exploration in the future.

\begin{acknowledgements}
We thank the referee for instructive suggestions and comments. This research has made use of data obtained by the High Energy Astrophysics Science Archive Research Center (HEASARC), which is an online service provided by NASA/GSFC. This work is supported by the National Key Research and Development Program of China (2021YFA0718500), the Natural Science Foundation of China (12273026, 11773050, 11833007, 12073023, U1931203, 11903024, 11773055, U1838203, and U1938114), Youth Innovation
Promotion Association of CAS (ID: 2020265), and the grant No. 2022CFB167.
\end{acknowledgements}

\bibliographystyle{aasjournal} 
\bibliography{ref}

\begin{thebibliography}{}
\expandafter\ifx\csname natexlab\endcsname\relax\def\natexlab#1{#1}\fi
\providecommand{\url}[1]{\href{#1}{#1}}
\providecommand{\dodoi}[1]{doi:~\href{http://doi.org/#1}{\nolinkurl{#1}}}
\providecommand{\doeprint}[1]{\href{http://ascl.net/#1}{\nolinkurl{http://ascl.net/#1}}}
\providecommand{\doarXiv}[1]{\href{https://arxiv.org/abs/#1}{\nolinkurl{https://arxiv.org/abs/#1}}}

\bibitem[{{Aramburu Sanchez} {et~al.}(2020){Aramburu Sanchez}, {Neilsen}, \&
  {Steiner}}]{ara2020}
{Aramburu Sanchez}, P., {Neilsen}, J., \& {Steiner}, J. 2020, in American
  Astronomical Society Meeting Abstracts, Vol. 235, American Astronomical
  Society Meeting Abstracts \#235, 369.14

\bibitem[{{Arnaud}(1996)}]{arn1996}
{Arnaud}, K.~A. 1996, in ASP Conf. Ser., Vol. 101, Astronomical Data Analysis
  Software and Systems V, ed. G.~H. {Jacoby} \& J.~{Barnes}, 17

\bibitem[{{Atri} {et~al.}(2020){Atri}, {Miller-Jones}, {Bahramian}, {Plotkin},
  {Deller}, {Jonker}, {Maccarone}, {Sivakoff}, {Soria}, {Altamirano},
  {Belloni}, {Fender}, {Koerding}, {Maitra}, {Markoff}, {Migliari}, {Russell},
  {Russell}, {Sarazin}, {Tetarenko}, \& {Tudose}}]{atr2020}
{Atri}, P., {Miller-Jones}, J.~C.~A., {Bahramian}, A., {et~al.} 2020, \mnras,
  493, L81, \dodoi{10.1093/mnrasl/slaa010}

\bibitem[{{Bambi} {et~al.}(2017){Bambi}, {C{\'a}rdenas-Avenda{\~n}o}, {Dauser},
  {Garc{\'\i}a}, \& {Nampalliwar}}]{bambi2017}
{Bambi}, C., {C{\'a}rdenas-Avenda{\~n}o}, A., {Dauser}, T., {Garc{\'\i}a},
  J.~A., \& {Nampalliwar}, S. 2017, \apj, 842, 76,
  \dodoi{10.3847/1538-4357/aa74c0}

\bibitem[{Beloborodov(1999)}]{bel1999}
Beloborodov, A.~M. 1999, The Astrophysical Journal, 510, L123,
  \dodoi{10.1086/311810}

\bibitem[{{Beri} {et~al.}(2019){Beri}, {Tetarenko}, {Bahramian}, {Altamirano},
  {Gandhi}, {Sivakoff}, {Degenaar}, {Middleton}, {Wijnands}, {Hern{\'a}ndez
  Santisteban}, \& {Paice}}]{beri2019}
{Beri}, A., {Tetarenko}, B.~E., {Bahramian}, A., {et~al.} 2019, \mnras, 485,
  3064, \dodoi{10.1093/mnras/stz616}

\bibitem[{{Bharali} {et~al.}(2019){Bharali}, {Chandra}, {Chauhan},
  {Garc{\'\i}a}, {Roy}, {Boettcher}, \& {Boruah}}]{bha2019}
{Bharali}, P., {Chandra}, S., {Chauhan}, J., {et~al.} 2019, \mnras, 487, 3150,
  \dodoi{10.1093/mnras/stz1492}

\bibitem[{{Brenneman} \& {Reynolds}(2006)}]{brenneman2006}
{Brenneman}, L.~W., \& {Reynolds}, C.~S. 2006, \apj, 652, 1028,
  \dodoi{10.1086/508146}

\bibitem[{{Buisson} {et~al.}(2019){Buisson}, {Fabian}, {Barret}, {F{\"u}rst},
  {Gandhi}, {Garc{\'\i}a}, {Kara}, {Madsen}, {Miller}, {Parker}, {Shaw},
  {Tomsick}, \& {Walton}}]{buisson2019}
{Buisson}, D.~J.~K., {Fabian}, A.~C., {Barret}, D., {et~al.} 2019, \mnras, 490,
  1350, \dodoi{10.1093/mnras/stz2681}

\bibitem[{{Casares} {et~al.}(2009){Casares}, {Orosz}, {Zurita}, {Shahbaz},
  {Corral-Santana}, {McClintock}, {Garcia}, {Mart{\'\i}nez-Pais}, {Charles},
  {Fender}, \& {Remillard}}]{cas2009}
{Casares}, J., {Orosz}, J.~A., {Zurita}, C., {et~al.} 2009, \apjs, 181, 238,
  \dodoi{10.1088/0067-0049/181/1/238}

\bibitem[{{Chakraborty} {et~al.}(2021){Chakraborty}, {Ratheesh},
  {Bhattacharyya}, {Tomsick}, {Tombesi}, {Fukumura}, \& {Jaisawal}}]{sudip2021}
{Chakraborty}, S., {Ratheesh}, A., {Bhattacharyya}, S., {et~al.} 2021, \mnras,
  508, 475, \dodoi{10.1093/mnras/stab2530}

\bibitem[{{Chand} {et~al.}(2020){Chand}, {Agrawal}, {Dewangan}, {Tripathi}, \&
  {Thakur}}]{chand2020}
{Chand}, S., {Agrawal}, V.~K., {Dewangan}, G.~C., {Tripathi}, P., \& {Thakur},
  P. 2020, \apj, 893, 142, \dodoi{10.3847/1538-4357/ab829a}

\bibitem[{{Chauhan} {et~al.}(2019){Chauhan}, {Miller-Jones}, {Anderson},
  {Raja}, {Bahramian}, {Hotan}, {Indermuehle}, {Whiting}, {Allison},
  {Anderson}, {Bunton}, {Koribalski}, \& {Mahony}}]{cha2019}
{Chauhan}, J., {Miller-Jones}, J.~C.~A., {Anderson}, G.~E., {et~al.} 2019,
  \mnras, 488, L129, \dodoi{10.1093/mnrasl/slz113}

\bibitem[{{Chauhan} {et~al.}(2021){Chauhan}, {Miller-Jones}, {Raja}, {Allison},
  {Jacob}, {Anderson}, {Carotenuto}, {Corbel}, {Fender}, {Hotan}, {Whiting},
  {Woudt}, {Koribalski}, \& {Mahony}}]{cha2021}
{Chauhan}, J., {Miller-Jones}, J.~C.~A., {Raja}, W., {et~al.} 2021, \mnras,
  501, L60, \dodoi{10.1093/mnrasl/slaa195}

\bibitem[{{Connors} {et~al.}(2021){Connors}, {Garc{\'\i}a}, {Tomsick}, {Hare},
  {Dauser}, {Grinberg}, {Steiner}, {Mastroserio}, {Sridhar}, {Fabian}, {Jiang},
  {Parker}, {Harrison}, \& {Kallman}}]{con2021}
{Connors}, R., {Garc{\'\i}a}, J., {Tomsick}, J., {et~al.} 2021, arXiv e-prints,
  arXiv:2101.06343.
\newblock \doarXiv{2101.06343}

\bibitem[{Curran(2014)}]{curran2014}
Curran, P.~A. 2014, ArXiv e-prints, 1411, arXiv:1411.3816.
\newblock \url{http://adsabs.harvard.edu/abs/2014arXiv1411.3816C}

\bibitem[{{Dauser} {et~al.}(2014){Dauser}, {Garc{\'{\i}}a}, {Parker}, {Fabian},
  \& {Wilms}}]{dau2014}
{Dauser}, T., {Garc{\'{\i}}a}, J., {Parker}, M.~L., {Fabian}, A.~C., \&
  {Wilms}, J. 2014, \mnras, 444, L100, \dodoi{10.1093/mnrasl/slu125}

\bibitem[{{Dauser} {et~al.}(2016){Dauser}, {Garc{\'\i}a}, {Walton}, {Eikmann},
  {Kallman}, {McClintock}, \& {Wilms}}]{dauser2016}
{Dauser}, T., {Garc{\'\i}a}, J., {Walton}, D.~J., {et~al.} 2016, \aap, 590,
  A76, \dodoi{10.1051/0004-6361/201628135}

\bibitem[{{De Marco} {et~al.}(2021){De Marco}, {Zdziarski}, {Ponti},
  {Migliori}, {Belloni}, {Segovia Otero}, {Dzie{\l}ak}, \& {Lai}}]{demarco2021}
{De Marco}, B., {Zdziarski}, A.~A., {Ponti}, G., {et~al.} 2021, \aap, 654, A14,
  \dodoi{10.1051/0004-6361/202140567}

\bibitem[{{De Marco} {et~al.}(2017){De Marco}, {Ponti}, {Petrucci}, {Clavel},
  {Corbel}, {Belmont}, {Chakravorty}, {Coriat}, {Drappeau}, {Ferreira},
  {Henri}, {Malzac}, {Rodriguez}, {Tomsick}, {Ursini}, \&
  {Zdziarski}}]{demarco2017}
{De Marco}, B., {Ponti}, G., {Petrucci}, P.~O., {et~al.} 2017, \mnras, 471,
  1475, \dodoi{10.1093/mnras/stx1649}

\bibitem[{{Diaz} {et~al.}(2020){Diaz}, {Ar{\'e}valo},
  {Hern{\'a}ndez-Garc{\'\i}a}, {Bassani}, {Malizia},
  {Gonz{\'a}lez-Mart{\'\i}n}, {Ricci}, {Matt}, {Stern}, {May}, {Zezas}, \&
  {Bauer}}]{dia2020}
{Diaz}, Y., {Ar{\'e}valo}, P., {Hern{\'a}ndez-Garc{\'\i}a}, L., {et~al.} 2020,
  \mnras, 496, 5399, \dodoi{10.1093/mnras/staa1762}

\bibitem[{{Done} {et~al.}(2007){Done}, {Gierli{\'n}ski}, \&
  {Kubota}}]{done2007}
{Done}, C., {Gierli{\'n}ski}, M., \& {Kubota}, A. 2007, \aapr, 15, 1,
  \dodoi{10.1007/s00159-007-0006-1}

\bibitem[{{Dong} {et~al.}(2022){Dong}, {Liu}, {Tuo}, {Steiner}, {Ge},
  {Garc{\'\i}a}, \& {Cao}}]{dong2022}
{Dong}, Y., {Liu}, Z., {Tuo}, Y., {et~al.} 2022, \mnras, 514, 1422,
  \dodoi{10.1093/mnras/stac1466}

\bibitem[{{Draghis} {et~al.}(2020){Draghis}, {Miller}, {Cackett}, {Kammoun},
  {Reynolds}, {Tomsick}, \& {Zoghbi}}]{dra2020}
{Draghis}, P.~A., {Miller}, J.~M., {Cackett}, E.~M., {et~al.} 2020, \apj, 900,
  78, \dodoi{10.3847/1538-4357/aba2ec}

\bibitem[{{El-Batal} {et~al.}(2016){El-Batal}, {Miller}, {Reynolds}, {Boggs},
  {Chistensen}, {Craig}, {Fuerst}, {Hailey}, {Harrison}, {Stern}, {Tomsick},
  {Walton}, \& {Zhang}}]{ei2016}
{El-Batal}, A.~M., {Miller}, J.~M., {Reynolds}, M.~T., {et~al.} 2016, \apjl,
  826, L12, \dodoi{10.3847/2041-8205/826/1/L12}

\bibitem[{{Esin} {et~al.}(1997){Esin}, {McClintock}, \& {Narayan}}]{esin1997}
{Esin}, A.~A., {McClintock}, J.~E., \& {Narayan}, R. 1997, \apj, 489, 865,
  \dodoi{10.1086/304829}

\bibitem[{{Ezhikode} {et~al.}(2020){Ezhikode}, {Dewangan}, {Misra}, \&
  {Philip}}]{ezh2020}
{Ezhikode}, S.~H., {Dewangan}, G.~C., {Misra}, R., \& {Philip}, N.~S. 2020,
  \mnras, 495, 3373, \dodoi{10.1093/mnras/staa1288}

\bibitem[{{Fabian} {et~al.}(2015){Fabian}, {Lohfink}, {Kara}, {Parker},
  {Vasudevan}, \& {Reynolds}}]{fabian2015}
{Fabian}, A.~C., {Lohfink}, A., {Kara}, E., {et~al.} 2015, \mnras, 451, 4375,
  \dodoi{10.1093/mnras/stv1218}

\bibitem[{{Feng} {et~al.}(2022){Feng}, {Zhao}, {Gou}, {Li}, {Steiner},
  {Garc{\'\i}a}, {Wang}, {Jia}, {Liao}, \& {Li}}]{feng2022}
{Feng}, Y., {Zhao}, X., {Gou}, L., {et~al.} 2022, Science China Physics,
  Mechanics, and Astronomy, 65, 219512, \dodoi{10.1007/s11433-021-1790-7}

\bibitem[{{Foreman-Mackey} {et~al.}(2013){Foreman-Mackey}, {Hogg}, {Lang}, \&
  {Goodman}}]{Foreman2013}
{Foreman-Mackey}, D., {Hogg}, D.~W., {Lang}, D., \& {Goodman}, J. 2013, \pasp,
  125, 306, \dodoi{10.1086/670067}

\bibitem[{{F{\"u}rst} {et~al.}(2015){F{\"u}rst}, {Nowak}, {Tomsick}, {Miller},
  {Corbel}, {Bachetti}, {Boggs}, {Christensen}, {Craig}, {Fabian}, {Gandhi},
  {Grinberg}, {Hailey}, {Harrison}, {Kara}, {Kennea}, {Madsen}, {Pottschmidt},
  {Stern}, {Walton}, {Wilms}, \& {Zhang}}]{furst2015}
{F{\"u}rst}, F., {Nowak}, M.~A., {Tomsick}, J.~A., {et~al.} 2015, \apj, 808,
  122, \dodoi{10.1088/0004-637X/808/2/122}

\bibitem[{{F{\"u}rst} {et~al.}(2016){F{\"u}rst}, {Tomsick}, {Yamaoka},
  {Dauser}, {Miller}, {Clavel}, {Corbel}, {Fabian}, {Garc{\'\i}a}, {Harrison},
  {Loh}, {Kaaret}, {Kalemci}, {Migliari}, {Miller-Jones}, {Pottschmidt},
  {Rahoui}, {Rodriguez}, {Stern}, {Stuhlinger}, {Walton}, \&
  {Wilms}}]{furst2016}
{F{\"u}rst}, F., {Tomsick}, J.~A., {Yamaoka}, K., {et~al.} 2016, \apj, 832,
  115, \dodoi{10.3847/0004-637X/832/2/115}

\bibitem[{{Gandhi} {et~al.}(2019){Gandhi}, {Rao}, {Johnson}, {Paice}, \&
  {Maccarone}}]{gan2019}
{Gandhi}, P., {Rao}, A., {Johnson}, M. A.~C., {Paice}, J.~A., \& {Maccarone},
  T.~J. 2019, \mnras, 485, 2642, \dodoi{10.1093/mnras/stz438}

\bibitem[{{Garc{\'\i}a} {et~al.}(2013){Garc{\'\i}a}, {Dauser}, {Reynolds},
  {Kallman}, {McClintock}, {Wilms}, \& {Eikmann}}]{gar2013}
{Garc{\'\i}a}, J., {Dauser}, T., {Reynolds}, C.~S., {et~al.} 2013, \apj, 768,
  146, \dodoi{10.1088/0004-637X/768/2/146}

\bibitem[{{Garc{\'{\i}}a} {et~al.}(2014){Garc{\'{\i}}a}, {Dauser}, {Lohfink},
  {Kallman}, {Steiner}, {McClintock}, {Brenneman}, {Wilms}, {Eikmann},
  {Reynolds}, \& {Tombesi}}]{gar2014}
{Garc{\'{\i}}a}, J., {Dauser}, T., {Lohfink}, A., {et~al.} 2014, \apj, 782, 76,
  \dodoi{10.1088/0004-637X/782/2/76}

\bibitem[{{Garc{\'\i}a} {et~al.}(2015{\natexlab{a}}){Garc{\'\i}a}, {Dauser},
  {Steiner}, {McClintock}, {Keck}, \& {Wilms}}]{garcia_dauser2015}
{Garc{\'\i}a}, J.~A., {Dauser}, T., {Steiner}, J.~F., {et~al.}
  2015{\natexlab{a}}, \apjl, 808, L37, \dodoi{10.1088/2041-8205/808/2/L37}

\bibitem[{{Garc{\'\i}a} {et~al.}(2015{\natexlab{b}}){Garc{\'\i}a}, {Steiner},
  {McClintock}, {Remillard}, {Grinberg}, \& {Dauser}}]{garcia2015}
{Garc{\'\i}a}, J.~A., {Steiner}, J.~F., {McClintock}, J.~E., {et~al.}
  2015{\natexlab{b}}, \apj, 813, 84, \dodoi{10.1088/0004-637X/813/2/84}

\bibitem[{{Garc{\'\i}a} {et~al.}(2019){Garc{\'\i}a}, {Tomsick}, {Sridhar},
  {Grinberg}, {Connors}, {Wang}, {Steiner}, {Dauser}, {Walton}, {Xu},
  {Harrison}, {Foster}, {Grefenstette}, {Madsen}, \& {Fabian}}]{garcia2019}
{Garc{\'\i}a}, J.~A., {Tomsick}, J.~A., {Sridhar}, N., {et~al.} 2019, \apj,
  885, 48, \dodoi{10.3847/1538-4357/ab384f}

\bibitem[{{George} \& {Fabian}(1991)}]{george1991}
{George}, I.~M., \& {Fabian}, A.~C. 1991, \mnras, 249, 352,
  \dodoi{10.1093/mnras/249.2.352}

\bibitem[{{Gilfanov} {et~al.}(1999){Gilfanov}, {Churazov}, \&
  {Revnivtsev}}]{gil1999}
{Gilfanov}, M., {Churazov}, E., \& {Revnivtsev}, M. 1999, \aap, 352, 182.
\newblock \doarXiv{astro-ph/9910084}

\bibitem[{{Harrison} {et~al.}(2013){Harrison}, {Craig}, {Christensen},
  {Hailey}, {Zhang}, {Boggs}, {Stern}, {Cook}, {Forster}, {Giommi},
  {Grefenstette}, {Kim}, {Kitaguchi}, {Koglin}, {Madsen}, {Mao}, {Miyasaka},
  {Mori}, {Perri}, {Pivovaroff}, {Puccetti}, {Rana}, {Westergaard}, {Willis},
  {Zoglauer}, {An}, {Bachetti}, {Barri{\`e}re}, {Bellm}, {Bhalerao},
  {Brejnholt}, {Fuerst}, {Liebe}, {Markwardt}, {Nynka}, {Vogel}, {Walton},
  {Wik}, {Alexander}, {Cominsky}, {Hornschemeier}, {Hornstrup}, {Kaspi},
  {Madejski}, {Matt}, {Molendi}, {Smith}, {Tomsick}, {Ajello}, {Ballantyne},
  {Balokovi{\'c}}, {Barret}, {Bauer}, {Blandford}, {Brandt}, {Brenneman},
  {Chiang}, {Chakrabarty}, {Chenevez}, {Comastri}, {Dufour}, {Elvis}, {Fabian},
  {Farrah}, {Fryer}, {Gotthelf}, {Grindlay}, {Helfand}, {Krivonos}, {Meier},
  {Miller}, {Natalucci}, {Ogle}, {Ofek}, {Ptak}, {Reynolds}, {Rigby},
  {Tagliaferri}, {Thorsett}, {Treister}, \& {Urry}}]{har2013}
{Harrison}, F.~A., {Craig}, W.~W., {Christensen}, F.~E., {et~al.} 2013, \apj,
  770, 103, \dodoi{10.1088/0004-637X/770/2/103}

\bibitem[{{Heida} {et~al.}(2017){Heida}, {Jonker}, {Torres}, \&
  {Chiavassa}}]{hei2017}
{Heida}, M., {Jonker}, P.~G., {Torres}, M.~A.~P., \& {Chiavassa}, A. 2017,
  \apj, 846, 132, \dodoi{10.3847/1538-4357/aa85df}

\bibitem[{{Ingram} {et~al.}(2016){Ingram}, {van der Klis}, {Middleton}, {Done},
  {Altamirano}, {Heil}, {Uttley}, \& {Axelsson}}]{ingram2016}
{Ingram}, A., {van der Klis}, M., {Middleton}, M., {et~al.} 2016, \mnras, 461,
  1967, \dodoi{10.1093/mnras/stw1245}

\bibitem[{{Iyer} {et~al.}(2015){Iyer}, {Nandi}, \& {Mandal}}]{iye2015}
{Iyer}, N., {Nandi}, A., \& {Mandal}, S. 2015, \apj, 807, 108,
  \dodoi{10.1088/0004-637X/807/1/108}

\bibitem[{{Jana} {et~al.}(2021){Jana}, {Jaisawal}, {Naik}, {Kumari},
  {Chatterjee}, {Chatterjee}, {Bhowmick}, {Chakrabarti}, {Chang}, \&
  {Debnath}}]{jan2021}
{Jana}, A., {Jaisawal}, G.~K., {Naik}, S., {et~al.} 2021, Research in Astronomy
  and Astrophysics, 21, 125, \dodoi{10.1088/1674-4527/21/5/125}

\bibitem[{{Jia} {et~al.}(2022){Jia}, {Zhao}, {Gou}, {Garc{\'\i}a}, {Liao},
  {Feng}, {Li}, {Wang}, {Li}, \& {Wu}}]{jia2022}
{Jia}, N., {Zhao}, X., {Gou}, L., {et~al.} 2022, \mnras, 511, 3125,
  \dodoi{10.1093/mnras/stac121}

\bibitem[{{Jiang} {et~al.}(2019){Jiang}, {Fabian}, {Wang}, {Walton},
  {Garc{\'\i}a}, {Parker}, {Steiner}, \& {Tomsick}}]{jiang2019}
{Jiang}, J., {Fabian}, A.~C., {Wang}, J., {et~al.} 2019, \mnras, 484, 1972,
  \dodoi{10.1093/mnras/stz095}

\bibitem[{{Jiang} {et~al.}(2020){Jiang}, {F{\"u}rst}, {Walton}, {Parker}, \&
  {Fabian}}]{jiang2020}
{Jiang}, J., {F{\"u}rst}, F., {Walton}, D.~J., {Parker}, M.~L., \& {Fabian},
  A.~C. 2020, \mnras, 492, 1947, \dodoi{10.1093/mnras/staa017}

\bibitem[{{Jiang} {et~al.}(2022){Jiang}, {Buisson}, {Dauser}, {Fabian},
  {F{\"u}rst}, {Gallo}, {Harrison}, {Parker}, {Steiner}, {Tomsick}, {Ubach}, \&
  {Walton}}]{jiang2022}
{Jiang}, J., {Buisson}, D. J.~K., {Dauser}, T., {et~al.} 2022, \mnras, 514,
  1952, \dodoi{10.1093/mnras/stac1401}

\bibitem[{{Kang} {et~al.}(2020){Kang}, {Wang}, \& {Kang}}]{kan2020}
{Kang}, J., {Wang}, J., \& {Kang}, W. 2020, \apj, 901, 111,
  \dodoi{10.3847/1538-4357/abadf5}

\bibitem[{{Kara} {et~al.}(2019){Kara}, {Steiner}, {Fabian}, {Cackett},
  {Uttley}, {Remillard}, {Gendreau}, {Arzoumanian}, {Altamirano}, {Eikenberry},
  {Enoto}, {Homan}, {Neilsen}, \& {Stevens}}]{kara2019}
{Kara}, E., {Steiner}, J.~F., {Fabian}, A.~C., {et~al.} 2019, \nat, 565, 198,
  \dodoi{10.1038/s41586-018-0803-x}

\bibitem[{{Kawamura} {et~al.}(2022){Kawamura}, {Axelsson}, {Done}, \&
  {Takahashi}}]{kawamura2022}
{Kawamura}, T., {Axelsson}, M., {Done}, C., \& {Takahashi}, T. 2022, \mnras,
  511, 536, \dodoi{10.1093/mnras/stac045}

\bibitem[{{Kelly}(2007)}]{kelly2007}
{Kelly}, B.~C. 2007, \apj, 665, 1489, \dodoi{10.1086/519947}

\bibitem[{{Klepczarek} {et~al.}(2023){Klepczarek}, {Nied{\'z}wiecki}, \&
  {Szanecki}}]{klepczarek2023}
{Klepczarek}, {\L}., {Nied{\'z}wiecki}, A., \& {Szanecki}, M. 2023, \mnras,
  519, L79, \dodoi{10.1093/mnrasl/slac156}

\bibitem[{{Koljonen} \& {Tomsick}(2020)}]{2020kol}
{Koljonen}, K.~I.~I., \& {Tomsick}, J.~A. 2020, \aap, 639, A13,
  \dodoi{10.1051/0004-6361/202037882}

\bibitem[{{Kubota} \& {Makishima}(2004)}]{kubota2004}
{Kubota}, A., \& {Makishima}, K. 2004, \apj, 601, 428, \dodoi{10.1086/380433}

\bibitem[{{Kubota} {et~al.}(1998){Kubota}, {Tanaka}, {Makishima}, {Ueda},
  {Dotani}, {Inoue}, \& {Yamaoka}}]{kubota1998}
{Kubota}, A., {Tanaka}, Y., {Makishima}, K., {et~al.} 1998, \pasj, 50, 667,
  \dodoi{10.1093/pasj/50.6.667}

\bibitem[{{Liu} \& {Qiao}(2022)}]{liu2022}
{Liu}, B.~F., \& {Qiao}, E. 2022, iSci, 25, 103544,
  \dodoi{10.1016/j.isci.2021.103544}

\bibitem[{{Madsen} {et~al.}(2020){Madsen}, {Grefenstette}, {Pike}, {Miyasaka},
  {Brightman}, {Forster}, \& {Harrison}}]{madsen2020}
{Madsen}, K.~K., {Grefenstette}, B.~W., {Pike}, S., {et~al.} 2020, arXiv
  e-prints, arXiv:2005.00569.
\newblock \doarXiv{2005.00569}

\bibitem[{{Mahmoud} \& {Done}(2018)}]{mahmoud2018}
{Mahmoud}, R.~D., \& {Done}, C. 2018, \mnras, 480, 4040,
  \dodoi{10.1093/mnras/sty2133}

\bibitem[{{Marino} {et~al.}(2021){Marino}, {Barnier}, {Petrucci}, {Del Santo},
  {Malzac}, {Ferreira}, {Marcel}, {Segreto}, {Motta}, {D'A{\`\i}}, {Di Salvo},
  {Guillot}, \& {Russell}}]{marino2021}
{Marino}, A., {Barnier}, S., {Petrucci}, P.~O., {et~al.} 2021, \aap, 656, A63,
  \dodoi{10.1051/0004-6361/202141146}

\bibitem[{{Miller} {et~al.}(2013){Miller}, {Parker}, {Fuerst}, {Bachetti},
  {Harrison}, {Barret}, {Boggs}, {Chakrabarty}, {Christensen}, {Craig},
  {Fabian}, {Grefenstette}, {Hailey}, {King}, {Stern}, {Tomsick}, {Walton}, \&
  {Zhang}}]{mil2013}
{Miller}, J.~M., {Parker}, M.~L., {Fuerst}, F., {et~al.} 2013, \apjl, 775, L45,
  \dodoi{10.1088/2041-8205/775/2/L45}

\bibitem[{Miller {et~al.}(2015)Miller, Tomsick, Bachetti, Wilkins, Boggs,
  Christensen, Craig, Fabian, Grefenstette, Hailey, Harrison, Kara, King,
  Stern, \& Zhang}]{mil2015}
Miller, J.~M., Tomsick, J.~A., Bachetti, M., {et~al.} 2015, The Astrophysical
  Journal, 799, L6, \dodoi{10.1088/2041-8205/799/1/l6}

\bibitem[{{Mu{\~n}oz-Darias} {et~al.}(2013){Mu{\~n}oz-Darias}, {Coriat},
  {Plant}, {Ponti}, {Fender}, \& {Dunn}}]{munoz2013}
{Mu{\~n}oz-Darias}, T., {Coriat}, M., {Plant}, D.~S., {et~al.} 2013, \mnras,
  432, 1330, \dodoi{10.1093/mnras/stt546}

\bibitem[{{Neustroev} {et~al.}(2014){Neustroev}, {Veledina}, {Poutanen},
  {Zharikov}, {Tsygankov}, {Sjoberg}, \& {Kajava}}]{neu2014}
{Neustroev}, V.~V., {Veledina}, A., {Poutanen}, J., {et~al.} 2014, \mnras, 445,
  2424, \dodoi{10.1093/mnras/stu1924}

\bibitem[{{Panagiotou} \& {Walter}(2020)}]{pan2020}
{Panagiotou}, C., \& {Walter}, R. 2020, \aap, 640, A31,
  \dodoi{10.1051/0004-6361/201937390}

\bibitem[{{Plant} {et~al.}(2014){Plant}, {Fender}, {Ponti}, {Mu{\~n}oz-Darias},
  \& {Coriat}}]{plant2014}
{Plant}, D.~S., {Fender}, R.~P., {Ponti}, G., {Mu{\~n}oz-Darias}, T., \&
  {Coriat}, M. 2014, \mnras, 442, 1767, \dodoi{10.1093/mnras/stu867}

\bibitem[{{Plant} {et~al.}(2015){Plant}, {Fender}, {Ponti}, {Mu{\~n}oz-Darias},
  \& {Coriat}}]{plant2015}
---. 2015, \aap, 573, A120, \dodoi{10.1051/0004-6361/201423925}

\bibitem[{{Ponti} {et~al.}(2012){Ponti}, {Fender}, {Begelman}, {Dunn},
  {Neilsen}, \& {Coriat}}]{ponti2012}
{Ponti}, G., {Fender}, R.~P., {Begelman}, M.~C., {et~al.} 2012, \mnras, 422,
  L11, \dodoi{10.1111/j.1745-3933.2012.01224.x}

\bibitem[{{Poutanen} {et~al.}(2018){Poutanen}, {Veledina}, \&
  {Zdziarski}}]{pou2018}
{Poutanen}, J., {Veledina}, A., \& {Zdziarski}, A.~A. 2018, \aap, 614, A79,
  \dodoi{10.1051/0004-6361/201732345}

\bibitem[{{Poutanen} {et~al.}(2022){Poutanen}, {Veledina}, {Berdyugin},
  {Berdyugina}, {Jermak}, {Jonker}, {Kajava}, {Kosenkov}, {Kravtsov},
  {Piirola}, {Shrestha}, {Perez Torres}, \& {Tsygankov}}]{pou2022}
{Poutanen}, J., {Veledina}, A., {Berdyugin}, A.~V., {et~al.} 2022, Science,
  375, 874, \dodoi{10.1126/science.abl4679}

\bibitem[{{Prabhakar} {et~al.}(2022){Prabhakar}, {Mandal}, {Athulya}, \&
  {Nandi}}]{prabhakar2022}
{Prabhakar}, G., {Mandal}, S., {Athulya}, M.~P., \& {Nandi}, A. 2022, \mnras,
  514, 6102, \dodoi{10.1093/mnras/stac1176}

\bibitem[{{Qiao} \& {Liu}(2012)}]{qiao2012}
{Qiao}, E., \& {Liu}, B.~F. 2012, \apj, 744, 145,
  \dodoi{10.1088/0004-637X/744/2/145}

\bibitem[{{Qiao} \& {Liu}(2017)}]{qia2017}
---. 2017, \mnras, 467, 898, \dodoi{10.1093/mnras/stx121}

\bibitem[{{Reid} {et~al.}(2014){Reid}, {McClintock}, {Steiner}, {Steeghs},
  {Remillard}, {Dhawan}, \& {Narayan}}]{rei2014}
{Reid}, M.~J., {McClintock}, J.~E., {Steiner}, J.~F., {et~al.} 2014, \apj, 796,
  2, \dodoi{10.1088/0004-637X/796/1/2}

\bibitem[{{Revnivtsev} {et~al.}(2001){Revnivtsev}, {Gilfanov}, \&
  {Churazov}}]{rev2001}
{Revnivtsev}, M., {Gilfanov}, M., \& {Churazov}, E. 2001, \aap, 380, 520,
  \dodoi{10.1051/0004-6361:20011413}

\bibitem[{{Ross} \& {Fabian}(2005)}]{ross2005}
{Ross}, R.~R., \& {Fabian}, A.~C. 2005, \mnras, 358, 211,
  \dodoi{10.1111/j.1365-2966.2005.08797.x}

\bibitem[{{Rout} {et~al.}(2022){Rout}, {Vadawale}, {Garcia}, \&
  {Connors}}]{rout2022}
{Rout}, S.~K., {Vadawale}, S., {Garcia}, J., \& {Connors}, R. 2022, arXiv
  e-prints, arXiv:2212.05293.
\newblock \doarXiv{2212.05293}

\bibitem[{{Saikia} {et~al.}(2022){Saikia}, {Russell}, {Baglio}, {Bramich},
  {Casella}, {Trigo}, {Gandhi}, {Jiang}, {Maccarone}, {Soria}, {Al Noori}, {Al
  Yazeedi}, {Alabarta}, {Belloni}, {Bel}, {Ceccobello}, {Corbel}, {Fender},
  {Gallo}, {Homan}, {Koljonen}, {Lewis}, {Markoff}, {Miller-Jones},
  {Rodriguez}, {Russell}, {Shahbaz}, {Sivakoff}, {Testa}, \&
  {Tetarenko}}]{saikia2022}
{Saikia}, P., {Russell}, D.~M., {Baglio}, M.~C., {et~al.} 2022, \apj, 932, 38,
  \dodoi{10.3847/1538-4357/ac6ce1}

\bibitem[{{Shaw} {et~al.}(2016{\natexlab{a}}){Shaw}, {Charles}, {Casares}, \&
  {Hern{\'a}ndez Santisteban}}]{sha2016}
{Shaw}, A.~W., {Charles}, P.~A., {Casares}, J., \& {Hern{\'a}ndez Santisteban},
  J.~V. 2016{\natexlab{a}}, \mnras, 463, 1314, \dodoi{10.1093/mnras/stw2092}

\bibitem[{{Shaw} {et~al.}(2016{\natexlab{b}}){Shaw}, {Gandhi}, {Altamirano},
  {Uttley}, {Tomsick}, {Charles}, {F{\"u}rst}, {Rahoui}, \&
  {Walton}}]{shaw2016}
{Shaw}, A.~W., {Gandhi}, P., {Altamirano}, D., {et~al.} 2016{\natexlab{b}},
  \mnras, 458, 1636, \dodoi{10.1093/mnras/stw417}

\bibitem[{{Shreeram} \& {Ingram}(2020)}]{shreeram2020}
{Shreeram}, S., \& {Ingram}, A. 2020, \mnras, 492, 405,
  \dodoi{10.1093/mnras/stz3455}

\bibitem[{{Sridhar} {et~al.}(2020){Sridhar}, {Garc{\'\i}a}, {Steiner},
  {Connors}, {Grinberg}, \& {Harrison}}]{sridhar2020}
{Sridhar}, N., {Garc{\'\i}a}, J.~A., {Steiner}, J.~F., {et~al.} 2020, \apj,
  890, 53, \dodoi{10.3847/1538-4357/ab64f5}

\bibitem[{{Steiner} {et~al.}(2012){Steiner}, {McClintock}, \& {Reid}}]{ste2012}
{Steiner}, J.~F., {McClintock}, J.~E., \& {Reid}, M.~J. 2012, \apjl, 745, L7,
  \dodoi{10.1088/2041-8205/745/1/L7}

\bibitem[{{Steiner} {et~al.}(2016){Steiner}, {Remillard}, {Garc{\'\i}a}, \&
  {McClintock}}]{ste2016}
{Steiner}, J.~F., {Remillard}, R.~A., {Garc{\'\i}a}, J.~A., \& {McClintock},
  J.~E. 2016, \apjl, 829, L22, \dodoi{10.3847/2041-8205/829/2/L22}

\bibitem[{{Stiele} \& {Kong}(2021)}]{stiele2021}
{Stiele}, H., \& {Kong}, A.~K.~H. 2021, \apj, 914, 93,
  \dodoi{10.3847/1538-4357/abfaa5}

\bibitem[{{Torres} {et~al.}(2020){Torres}, {Casares}, {Jim{\'e}nez-Ibarra},
  {{\'A}lvarez-Hern{\'a}ndez}, {Mu{\~n}oz-Darias}, {Armas Padilla}, {Jonker},
  \& {Heida}}]{torres2020}
{Torres}, M.~A.~P., {Casares}, J., {Jim{\'e}nez-Ibarra}, F., {et~al.} 2020,
  \apjl, 893, L37, \dodoi{10.3847/2041-8213/ab863a}

\bibitem[{{Tripathi} {et~al.}(2021){Tripathi}, {Zhang}, {Abdikamalov},
  {Ayzenberg}, {Bambi}, {Jiang}, {Liu}, \& {Zhou}}]{tripathi2021}
{Tripathi}, A., {Zhang}, Y., {Abdikamalov}, A.~B., {et~al.} 2021, \apj, 913,
  79, \dodoi{10.3847/1538-4357/abf6cd}

\bibitem[{{Ueda} {et~al.}(1994){Ueda}, {Ebisawa}, \& {Done}}]{ued1994}
{Ueda}, Y., {Ebisawa}, K., \& {Done}, C. 1994, \pasj, 46, 107

\bibitem[{{Uttley} {et~al.}(2018){Uttley}, {Gendreau}, {Markwardt},
  {Strohmayer}, {Bult}, {Arzoumanian}, {Pottschmidt}, {Ray}, {Remillard},
  {Pasham}, {Steiner}, {Neilsen}, {Homan}, {Miller}, {Iwakiri}, \&
  {Fabian}}]{utt2018}
{Uttley}, P., {Gendreau}, K., {Markwardt}, C., {et~al.} 2018, The Astronomer's
  Telegram, 11423, 1

\bibitem[{{Walton} {et~al.}(2017){Walton}, {Mooley}, {King}, {Tomsick},
  {Miller}, {Dauser}, {Garc{\'\i}a}, {Bachetti}, {Brightman}, {Fabian},
  {Forster}, {F{\"u}rst}, {Gandhi}, {Grefenstette}, {Harrison}, {Madsen},
  {Meier}, {Middleton}, {Natalucci}, {Rahoui}, {Rana}, \& {Stern}}]{walton2017}
{Walton}, D.~J., {Mooley}, K., {King}, A.~L., {et~al.} 2017, \apj, 839, 110,
  \dodoi{10.3847/1538-4357/aa67e8}

\bibitem[{{Wang} {et~al.}(2020{\natexlab{a}}){Wang}, {Kara}, {Steiner},
  {Garc{\'\i}a}, {Homan}, {Neilsen}, {Marcel}, {Ludlam}, {Tombesi}, {Cackett},
  \& {Remillard}}]{wangj2020}
{Wang}, J., {Kara}, E., {Steiner}, J.~F., {et~al.} 2020{\natexlab{a}}, \apj,
  899, 44, \dodoi{10.3847/1538-4357/ab9ec3}

\bibitem[{{Wang} {et~al.}(2022){Wang}, {Kong}, {Chen}, {Zhang}, {Zhang},
  {Soria}, {Ji}, {Qu}, {Huang}, {Tao}, {Ge}, {Lu}, {Chen}, {Li}, {Xu}, {Cao},
  {Chen}, {Liu}, {Bu}, {Cai}, {Chang}, {Chen}, {Chen}, {Cui}, {Du}, {Gao},
  {Gao}, {Gu}, {Guan}, {Guo}, {Han}, {Huo}, {Jia}, {Jiang}, {Jin}, {Li}, {Li},
  {Li}, {Li}, {Li}, {Li}, {Li}, {Li}, {Liang}, {Liao}, {Liu}, {Liu}, {Liu},
  {Liu}, {Lu}, {Luo}, {Luo}, {Ma}, {Ma}, {Meng}, {Nang}, {Nie}, {Ou}, {Ren},
  {Sai}, {Song}, {Song}, {Sun}, {Tan}, {Tuo}, {Wang}, {Wang}, {Wang}, {Wang},
  {Wen}, {Wu}, {Wu}, {Wu}, {Xiao}, {Xiao}, {Xiong}, {Yang}, {Yang}, {Yang},
  {Yang}, {Yi}, {Yin}, {You}, {Zhang}, {Zhang}, {Zhang}, {Zhang}, {Zhang},
  {Zhang}, {Zhang}, {Zhang}, {Zhao}, {Zhao}, {Zheng}, {Zheng}, \&
  {Zhou}}]{wangpj2022}
{Wang}, P.~J., {Kong}, L.~D., {Chen}, Y.~P., {et~al.} 2022, \mnras, 512, 4541,
  \dodoi{10.1093/mnras/stac773}

\bibitem[{{Wang} {et~al.}(2020{\natexlab{b}}){Wang}, {Ji}, {Zhang},
  {M{\'e}ndez}, {Qu}, {Maggi}, {Ge}, {Qiao}, {Tao}, {Zhang}, {Altamirano},
  {Zhang}, {Ma}, {Lu}, {Li}, {Huang}, {Zheng}, {Chen}, {Chang}, {Tuo},
  {G{\"u}ng{\"o}r}, {Song}, {Xu}, {Cao}, {Chen}, {Liu}, {Bu}, {Cai}, {Chen},
  {Chen}, {Chen}, {Chen}, {Cui}, {Cui}, {Deng}, {Dong}, {Du}, {Fu}, {Gao},
  {Gao}, {Gao}, {Gu}, {Guan}, {Guo}, {Han}, {Huo}, {Jia}, {Jiang}, {Jiang},
  {Jin}, {Jin}, {Kong}, {Li}, {Li}, {Li}, {Li}, {Li}, {Li}, {Li}, {Li}, {Li},
  {Li}, {Liang}, {Liao}, {Liu}, {Liu}, {Liu}, {Liu}, {Lu}, {Lu}, {Luo}, {Luo},
  {Meng}, {Nang}, {Nie}, {Ou}, {Sai}, {Shang}, {Song}, {Sun}, {Tan}, {Wang},
  {Wang}, {Wang}, {Wang}, {Wang}, {Wen}, {Wu}, {Wu}, {Wu}, {Xiao}, {Xiao},
  {Xiong}, {Yang}, {Yang}, {Yang}, {Yang}, {Yi}, {Yin}, {You}, {Zhang},
  {Zhang}, {Zhang}, {Zhang}, {Zhang}, {Zhang}, {Zhang}, {Zhang}, {Zhang},
  {Zhang}, {Zhang}, {Zhang}, {Zhang}, {Zhang}, {Zhang}, {Zhang}, {Zhao},
  {Zhao}, {Zhou}, {Zhou}, {Zhuang}, {Zhu}, {Zhu}, \& {Wang}}]{wang2020}
{Wang}, Y., {Ji}, L., {Zhang}, S.~N., {et~al.} 2020{\natexlab{b}}, \apj, 896,
  33, \dodoi{10.3847/1538-4357/ab8db4}

\bibitem[{{Wang-Ji} {et~al.}(2018){Wang-Ji}, {Garc{\'\i}a}, {Steiner},
  {Tomsick}, {Harrison}, {Bambi}, {Petrucci}, {Ferreira}, {Chakravorty}, \&
  {Clavel}}]{wj2018}
{Wang-Ji}, J., {Garc{\'\i}a}, J.~A., {Steiner}, J.~F., {et~al.} 2018, \apj,
  855, 61, \dodoi{10.3847/1538-4357/aaa974}

\bibitem[{{Wilms} {et~al.}(2000){Wilms}, {Allen}, \& {McCray}}]{wilms2000}
{Wilms}, J., {Allen}, A., \& {McCray}, R. 2000, \apj, 542, 914,
  \dodoi{10.1086/317016}

\bibitem[{{Wu} \& {Gu}(2008)}]{wu2008}
{Wu}, Q., \& {Gu}, M. 2008, \apj, 682, 212, \dodoi{10.1086/588187}

\bibitem[{{Xu} {et~al.}(2018{\natexlab{a}}){Xu}, {Nampalliwar}, {Abdikamalov},
  {Ayzenberg}, {Bambi}, {Dauser}, {Garc{\'\i}a}, \& {Jiang}}]{xu2018}
{Xu}, Y., {Nampalliwar}, S., {Abdikamalov}, A.~B., {et~al.} 2018{\natexlab{a}},
  \apj, 865, 134, \dodoi{10.3847/1538-4357/aadb9d}

\bibitem[{{Xu} {et~al.}(2017){Xu}, {Garc{\'\i}a}, {F{\"u}rst}, {Harrison},
  {Walton}, {Tomsick}, {Bachetti}, {King}, {Madsen}, {Miller}, \&
  {Grinberg}}]{xu2017}
{Xu}, Y., {Garc{\'\i}a}, J.~A., {F{\"u}rst}, F., {et~al.} 2017, \apj, 851, 103,
  \dodoi{10.3847/1538-4357/aa9ab4}

\bibitem[{{Xu} {et~al.}(2018{\natexlab{b}}){Xu}, {Harrison}, {Garc{\'\i}a},
  {Fabian}, {F{\"u}rst}, {Gandhi}, {Grefenstette}, {Madsen}, {Miller},
  {Parker}, {Tomsick}, \& {Walton}}]{xuyj2018}
{Xu}, Y., {Harrison}, F.~A., {Garc{\'\i}a}, J.~A., {et~al.} 2018{\natexlab{b}},
  \apjl, 852, L34, \dodoi{10.3847/2041-8213/aaa4b2}

\bibitem[{{Yamaoka} {et~al.}(2005){Yamaoka}, {Uzawa}, {Arai}, {Yamazaki}, \&
  {Yoshida}}]{yamaoka2005}
{Yamaoka}, K., {Uzawa}, M., {Arai}, M., {Yamazaki}, T., \& {Yoshida}, A. 2005,
  Chinese Journal of Astronomy and Astrophysics Supplement, 5, 273

\bibitem[{{Yan} {et~al.}(2020){Yan}, {Xie}, \& {Zhang}}]{yan2020}
{Yan}, Z., {Xie}, F.-G., \& {Zhang}, W. 2020, \apjl, 889, L18,
  \dodoi{10.3847/2041-8213/ab665e}

\bibitem[{{Yan} \& {Yu}(2017)}]{yan2017}
{Yan}, Z., \& {Yu}, W. 2017, \mnras, 470, 4298, \dodoi{10.1093/mnras/stx1562}

\bibitem[{{Yang} {et~al.}(2015){Yang}, {Xie}, {Yuan}, {Zdziarski},
  {Gierli{\'n}ski}, {Ho}, \& {Yu}}]{yang2015}
{Yang}, Q.-X., {Xie}, F.-G., {Yuan}, F., {et~al.} 2015, \mnras, 447, 1692,
  \dodoi{10.1093/mnras/stu2571}

\bibitem[{{You} {et~al.}(2012){You}, {Cao}, \& {Yuan}}]{you2012}
{You}, B., {Cao}, X., \& {Yuan}, Y.-F. 2012, \apj, 761, 109,
  \dodoi{10.1088/0004-637X/761/2/109}

\bibitem[{{You} {et~al.}(2021){You}, {Tuo}, {Li}, {Wang}, {Zhang}, {Zhang},
  {Ge}, {Luo}, {Liu}, {Yuan}, {Dai}, {Liu}, {Qiao}, {Jin}, {Liu}, {Czerny},
  {Wu}, {Bu}, {Cai}, {Cao}, {Chang}, {Chen}, {Chen}, {Chen}, {Chen}, {Chen},
  {Chen}, {Cui}, {Cui}, {Deng}, {Dong}, {Du}, {Fu}, {Gao}, {Gao}, {Gao}, {Gu},
  {Guan}, {Guo}, {Han}, {Huang}, {Huo}, {Jia}, {Jiang}, {Jiang}, {Jin}, {Jin},
  {Kong}, {Li}, {Li}, {Li}, {Li}, {Li}, {Li}, {Li}, {Li}, {Li}, {Li}, {Li},
  {Liang}, {Liao}, {Liu}, {Liu}, {Liu}, {Liu}, {Liu}, {Lu}, {Lu}, {Lu}, {Luo},
  {Luo}, {Ma}, {Meng}, {Nang}, {Nie}, {Ou}, {Qu}, {Sai}, {Shang}, {Song},
  {Song}, {Sun}, {Tan}, {Tao}, {Wang}, {Wang}, {Wang}, {Wang}, {Wang}, {Wang},
  {Wen}, {Wu}, {Wu}, {Wu}, {Xiao}, {Xiao}, {Xiong}, {Xu}, {Yang}, {Yang},
  {Yang}, {Yi}, {Yin}, {You}, {Zhang}, {Zhang}, {Zhang}, {Zhang}, {Zhang},
  {Zhang}, {Zhang}, {Zhang}, {Zhang}, {Zhang}, {Zhang}, {Zhang}, {Zhang},
  {Zhang}, {Zhang}, {Zhao}, {Zhao}, {Zheng}, {Zhou}, {Zhou}, {Zhu}, \&
  {Zhu}}]{you2021}
{You}, B., {Tuo}, Y., {Li}, C., {et~al.} 2021, Nature Communications, 12, 1025,
  \dodoi{10.1038/s41467-021-21169-5}

\bibitem[{{Yuan} \& {Narayan}(2014)}]{yuan2014}
{Yuan}, F., \& {Narayan}, R. 2014, \araa, 52, 529,
  \dodoi{10.1146/annurev-astro-082812-141003}

\bibitem[{{Zdziarski} {et~al.}(2021{\natexlab{a}}){Zdziarski}, {Dzie{\l}ak},
  {De Marco}, {Szanecki}, \& {Nied{\'z}wiecki}}]{aaz2021a}
{Zdziarski}, A.~A., {Dzie{\l}ak}, M.~A., {De Marco}, B., {Szanecki}, M., \&
  {Nied{\'z}wiecki}, A. 2021{\natexlab{a}}, \apjl, 909, L9,
  \dodoi{10.3847/2041-8213/abe7ef}

\bibitem[{{Zdziarski} {et~al.}(1996){Zdziarski}, {Johnson}, \&
  {Magdziarz}}]{aaz1996}
{Zdziarski}, A.~A., {Johnson}, W.~N., \& {Magdziarz}, P. 1996, \mnras, 283,
  193, \dodoi{10.1093/mnras/283.1.193}

\bibitem[{{Zdziarski} {et~al.}(1999){Zdziarski}, {Lubi{\'n}ski}, \&
  {Smith}}]{zdz1999}
{Zdziarski}, A.~A., {Lubi{\'n}ski}, P., \& {Smith}, D.~A. 1999, \mnras, 303,
  L11, \dodoi{10.1046/j.1365-8711.1999.02343.x}

\bibitem[{{Zdziarski} {et~al.}(2022{\natexlab{a}}){Zdziarski}, {You}, \&
  {Szanecki}}]{aaz2022}
{Zdziarski}, A.~A., {You}, B., \& {Szanecki}, M. 2022{\natexlab{a}}, \apjl,
  939, L2, \dodoi{10.3847/2041-8213/ac9474}

\bibitem[{{Zdziarski} {et~al.}(2022{\natexlab{b}}){Zdziarski}, {You},
  {Szanecki}, {Li}, \& {Ge}}]{aazyb2022}
{Zdziarski}, A.~A., {You}, B., {Szanecki}, M., {Li}, X.-B., \& {Ge}, M.
  2022{\natexlab{b}}, \apj, 928, 11, \dodoi{10.3847/1538-4357/ac54a7}

\bibitem[{{Zdziarski} {et~al.}(2019){Zdziarski}, {Zi{\'o}{\l}kowski}, \&
  {Miko{\l}ajewska}}]{aaz2019}
{Zdziarski}, A.~A., {Zi{\'o}{\l}kowski}, J., \& {Miko{\l}ajewska}, J. 2019,
  \mnras, 488, 1026, \dodoi{10.1093/mnras/stz1787}

\bibitem[{{Zdziarski} {et~al.}(2021{\natexlab{b}}){Zdziarski}, {Jourdain},
  {Lubi{\'n}ski}, {Szanecki}, {Nied{\'z}wiecki}, {Veledina}, {Poutanen},
  {Dzie{\l}ak}, \& {Roques}}]{aaz2021b}
{Zdziarski}, A.~A., {Jourdain}, E., {Lubi{\'n}ski}, P., {et~al.}
  2021{\natexlab{b}}, \apjl, 914, L5, \dodoi{10.3847/2041-8213/ac0147}

\bibitem[{{Zhang} {et~al.}(2019{\natexlab{a}}){Zhang}, {Abdikamalov},
  {Ayzenberg}, {Bambi}, {Dauser}, {Garc{\'\i}a}, \& {Nampalliwar}}]{zhanga2019}
{Zhang}, Y., {Abdikamalov}, A.~B., {Ayzenberg}, D., {et~al.}
  2019{\natexlab{a}}, \apj, 875, 41, \dodoi{10.3847/1538-4357/ab0e79}

\bibitem[{{Zhang} {et~al.}(2019{\natexlab{b}}){Zhang}, {Abdikamalov},
  {Ayzenberg}, {Bambi}, \& {Nampalliwar}}]{zhangb2019}
{Zhang}, Y., {Abdikamalov}, A.~B., {Ayzenberg}, D., {Bambi}, C., \&
  {Nampalliwar}, S. 2019{\natexlab{b}}, \apj, 884, 147,
  \dodoi{10.3847/1538-4357/ab4271}

\end{thebibliography}


\begin{longtable}{lcccccc}
\caption{\emph{NuSTAR} observation Details}\label{tab:obs_detail}\\

\hline
Source & ObsID & Detector & MJD & Exp. & Counts (s$^{-1}$)     & HR \\
\hline    
\endfirsthead
\caption{continued.}\\
\hline\hline
Source & ObsID & Detector & MJD & Exp. & Counts (s$^{-1}$)     & HR \\
 \hline
\endhead
\hline
\endfoot
GRS 1716-249 & 80201034006 & FPMA & 57758.50 & 6046.56 & 113.90 $\pm$ 0.14 & 0.47 \\
  &   & FPMB & 57758.50 & 6134.49 & 108.20 $\pm$ 0.13 & 0.46 \\
  & 80201034007 & FPMA & 57781.32 & 42465.19 & 129.50 $\pm$ 0.06 & 0.48 \\
  &   & FPMB & 57781.32 & 43515.15 & 119.30 $\pm$ 0.05 & 0.47 \\
  & 90202055002 & FPMA & 57850.61 & 17897.52 & 125.40 $\pm$ 0.08 & 0.40 \\
  &   & FPMB & 57850.61 & 18259.78 & 115.70 $\pm$ 0.08 & 0.39 \\
  & 90202055004 & FPMA & 57853.70 & 15797.47 & 121.70 $\pm$ 0.09 & 0.38 \\
  &   & FPMB & 57853.70 & 16100.69 & 112.70 $\pm$ 0.08 & 0.36 \\
\hline               
GRS 1739-278 & 80101050002 & FPMA & 57280.89 & 41241.62 & 0.35 $\pm$ 0.00 & 0.50 \\
  &   & FPMB & 57280.89 & 40965.48 & 0.38 $\pm$ 0.00 & 0.43 \\
  & 80102101002 & FPMA & 57660.89 & 26310.03 & 3.14 $\pm$ 0.01 & 0.49 \\
  &   & FPMB & 57660.89 & 26102.79 & 3.00 $\pm$ 0.01 & 0.47 \\
  & 80102101004 & FPMA & 57680.63 & 25897.14 & 0.88 $\pm$ 0.01 & 0.49 \\
  &   & FPMB & 57680.63 & 25703.41 & 0.88 $\pm$ 0.01 & 0.46 \\
  & 80102101005 & FPMA & 57692.85 & 27021.16 & 1.93 $\pm$ 0.01 & 0.49 \\
  &   & FPMB & 57692.85 & 26903.39 & 1.78 $\pm$ 0.01 & 0.47 \\
\hline               
GRS 1915+105 & 10002004001 & FPMA & 56111.06 & 14696.17 & 195.00 $\pm$ 0.12 & 0.32 \\
  &   & FPMB & 56111.06 & 15114.49 & 181.70 $\pm$ 0.11 & 0.32 \\
  & 80401312002 & FPMA & 58277.51 & 26167.74 & 82.96 $\pm$ 0.06 & 0.33 \\
  &   & FPMB & 58277.51 & 26396.31 & 76.44 $\pm$ 0.05 & 0.32 \\
  & 90202045002 & FPMA & 57806.17 & 13903.36 & 169.20 $\pm$ 0.11 & 0.31 \\
  &   & FPMB & 57806.17 & 14093.44 & 158.60 $\pm$ 0.11 & 0.30 \\
\hline               
GS 1354-64 & 90101006002 & FPMA & 57186.30 & 24021.18 & 6.94 $\pm$ 0.02 & 0.60 \\
  &   & FPMB & 57186.30 & 23694.70 & 6.37 $\pm$ 0.02 & 0.58 \\
  & 90101006004 & FPMA & 57214.59 & 28797.52 & 48.24 $\pm$ 0.04 & 0.52 \\
  &   & FPMB & 57214.59 & 28624.73 & 44.72 $\pm$ 0.04 & 0.50 \\
  & 90101006006 & FPMA & 57240.31 & 34965.82 & 54.92 $\pm$ 0.04 & 0.47 \\
  &   & FPMB & 57240.31 & 35030.46 & 51.89 $\pm$ 0.04 & 0.46 \\
\hline               
GX 339-4 & 80001013002 & FPMA & 56516.01 & 35090.43 & 10.41 $\pm$ 0.02 & 0.53 \\
  &   & FPMB & 56516.01 & 35098.32 & 10.15 $\pm$ 0.02 & 0.52 \\
  & 80001013004 & FPMA & 56520.75 & 20794.25 & 17.41 $\pm$ 0.03 & 0.53 \\
  &   & FPMB & 56520.75 & 20834.18 & 16.42 $\pm$ 0.03 & 0.52 \\
  & 80001013006 & FPMA & 56528.54 & 23152.06 & 27.74 $\pm$ 0.03 & 0.52 \\
  &   & FPMB & 56528.54 & 23244.36 & 26.10 $\pm$ 0.03 & 0.51 \\
  & 80001013008 & FPMA & 56538.44 & 43189.25 & 36.39 $\pm$ 0.03 & 0.51 \\
  &   & FPMB & 56538.44 & 43438.82 & 34.54 $\pm$ 0.03 & 0.50 \\
  & 80001013010 & FPMA & 56582.00 & 97569.81 & 6.16 $\pm$ 0.01 & 0.50 \\
  &   & FPMB & 56582.00 & 97944.22 & 5.87 $\pm$ 0.01 & 0.49 \\
  & 80001015001 & FPMA & 57035.24 & 10377.22 & 103.70 $\pm$ 0.10 & 0.38 \\
  &   & FPMB & 57035.24 & 10549.82 & 96.38 $\pm$ 0.10 & 0.37 \\
  & 80102011002 & FPMA & 57262.56 & 18249.12 & 18.76 $\pm$ 0.03 & 0.39 \\
  &   & FPMB & 57262.56 & 17658.19 & 17.54 $\pm$ 0.03 & 0.38 \\
  & 80102011004 & FPMA & 57267.53 & 17370.83 & 16.44 $\pm$ 0.03 & 0.45 \\
  &   & FPMB & 57267.53 & 17202.23 & 15.33 $\pm$ 0.03 & 0.44 \\
  & 80102011006 & FPMA & 57272.63 & 16239.13 & 13.87 $\pm$ 0.03 & 0.47 \\
  &   & FPMB & 57272.63 & 16091.85 & 13.03 $\pm$ 0.03 & 0.45 \\
  & 80102011010 & FPMA & 57282.43 & 35689.01 & 7.70 $\pm$ 0.01 & 0.49 \\
  &   & FPMB & 57282.43 & 35563.32 & 7.27 $\pm$ 0.01 & 0.47 \\
  & 80102011012 & FPMA & 57295.06 & 41332.23 & 4.15 $\pm$ 0.01 & 0.48 \\
  &   & FPMB & 57295.06 & 41322.83 & 3.87 $\pm$ 0.01 & 0.46 \\
  & 80302304002 & FPMA & 58028.15 & 20523.68 & 2.38 $\pm$ 0.01 & 0.48 \\
  &   & FPMB & 58028.15 & 20767.70 & 2.23 $\pm$ 0.01 & 0.46 \\
  & 80302304004 & FPMA & 58051.58 & 17046.56 & 21.97 $\pm$ 0.04 & 0.51 \\
  &   & FPMB & 58051.58 & 16956.34 & 20.58 $\pm$ 0.04 & 0.50 \\
  & 80302304005 & FPMA & 58059.90 & 18515.38 & 19.72 $\pm$ 0.03 & 0.51 \\
  &   & FPMB & 58059.90 & 18534.54 & 18.39 $\pm$ 0.03 & 0.49 \\
  & 80502325008 & FPMA & 58959.06 & 22397.06 & 10.33 $\pm$ 0.02 & 0.44 \\
  &   & FPMB & 58959.06 & 22166.46 & 9.57 $\pm$ 0.02 & 0.43 \\
  & 90401369004 & FPMA & 58488.63 & 3621.66 & 13.05 $\pm$ 0.06 & 0.52 \\
  &   & FPMB & 58488.63 & 3616.51 & 12.04 $\pm$ 0.06 & 0.50 \\
\hline               
H1743-322 & 80001044002 & FPMA & 56918.66 & 50252.45 & 26.22 $\pm$ 0.02 & 0.57 \\
  &   & FPMB & 56918.66 & 50387.07 & 25.39 $\pm$ 0.02 & 0.55 \\
  & 80001044004 & FPMA & 56923.77 & 61278.07 & 34.86 $\pm$ 0.02 & 0.54 \\
  &   & FPMB & 56923.77 & 61339.79 & 33.88 $\pm$ 0.02 & 0.53 \\
  & 80001044006 & FPMA & 56939.78 & 25666.98 & 31.02 $\pm$ 0.03 & 0.55 \\
  &   & FPMB & 56939.78 & 25721.70 & 29.17 $\pm$ 0.03 & 0.53 \\
  & 80002040002 & FPMA & 57206.13 & 27418.52 & 18.02 $\pm$ 0.03 & 0.56 \\
  &   & FPMB & 57206.13 & 27603.63 & 16.53 $\pm$ 0.02 & 0.54 \\
  & 80202012002 & FPMA & 57460.07 & 65869.27 & 38.61 $\pm$ 0.02 & 0.44 \\
  &   & FPMB & 57460.07 & 66283.17 & 36.43 $\pm$ 0.02 & 0.43 \\
  & 80202012004 & FPMA & 57462.29 & 64870.79 & 37.00 $\pm$ 0.02 & 0.44 \\
  &   & FPMB & 57462.29 & 65040.59 & 34.68 $\pm$ 0.02 & 0.43 \\
  & 80202012006 & FPMA & 58387.37 & 65702.67 & 20.10 $\pm$ 0.02 & 0.54 \\
  &   & FPMB & 58387.37 & 65610.73 & 18.99 $\pm$ 0.02 & 0.53 \\
  & 90401335002 & FPMA & 58380.12 & 38443.65 & 32.36 $\pm$ 0.03 & 0.52 \\
  &   & FPMB & 58380.12 & 38425.23 & 30.52 $\pm$ 0.03 & 0.51 \\
\hline               
IGR J17091-3624 & 80001041002 & FPMA & 57454.09 & 43297.76 & 15.80 $\pm$ 0.02 & 0.52 \\
  &   & FPMB & 57454.09 & 42957.81 & 14.75 $\pm$ 0.02 & 0.51 \\
  & 80202014002 & FPMA & 57459.60 & 20239.92 & 20.64 $\pm$ 0.03 & 0.50 \\
  &   & FPMB & 57459.60 & 20278.74 & 19.23 $\pm$ 0.03 & 0.48 \\
  & 80202014004 & FPMA & 57461.81 & 20700.15 & 22.79 $\pm$ 0.03 & 0.48 \\
  &   & FPMB & 57461.81 & 20515.16 & 21.24 $\pm$ 0.03 & 0.46 \\
\hline               
MAXI J1348-630 & 80402315002 & FPMA & 58515.13 & 3038.04 & 1043.00 $\pm$ 0.59 & 0.43 \\
  &   & FPMB & 58515.13 & 3208.37 & 906.00 $\pm$ 0.53 & 0.42 \\
  & 80402315004 & FPMA & 58515.60 & 736.34 & 1048.00 $\pm$ 1.20 & 0.42 \\
  &   & FPMB & 58515.60 & 778.54 & 911.50 $\pm$ 1.09 & 0.42 \\
  & 80502304002 & FPMA & 58655.40 & 13783.86 & 186.80 $\pm$ 0.12 & 0.45 \\
  &   & FPMB & 58655.40 & 14172.58 & 172.20 $\pm$ 0.11 & 0.44 \\
  & 80502304004 & FPMA & 58660.50 & 15367.24 & 161.90 $\pm$ 0.10 & 0.47 \\
  &   & FPMB & 58660.50 & 15773.25 & 149.50 $\pm$ 0.10 & 0.46 \\
  & 80502304006 & FPMA & 58672.38 & 17181.65 & 107.60 $\pm$ 0.08 & 0.48 \\
  &   & FPMB & 58672.38 & 17460.58 & 100.00 $\pm$ 0.08 & 0.47 \\
\hline               
MAXI J1535-571 & 90301013002 & FPMA & 58003.79 & 10142.25 & 609.40 $\pm$ 0.25 & 0.41 \\
  &   & FPMB & 58003.79 & 10670.31 & 549.40 $\pm$ 0.23 & 0.40 \\
\hline               
MAXI J1813-095 & 80402303002 & FPMA & 58177.50 & 20546.62 & 16.18 $\pm$ 0.03 & 0.50 \\
  &   & FPMB & 58177.50 & 20002.84 & 15.16 $\pm$ 0.03 & 0.49 \\
  & 80402303004 & FPMA & 58183.60 & 20433.92 & 13.43 $\pm$ 0.03 & 0.49 \\
  &   & FPMB & 58183.60 & 20346.84 & 12.69 $\pm$ 0.03 & 0.49 \\
  & 80402303006 & FPMA & 58202.73 & 23238.40 & 14.52 $\pm$ 0.03 & 0.46 \\
  &   & FPMB & 58202.73 & 23099.06 & 13.66 $\pm$ 0.02 & 0.45 \\
\hline               
MAXI J1820+070 & 90401309002 & FPMA & 58191.94 & 11769.40 & 178.90 $\pm$ 0.12 & 0.56 \\
  &   & FPMB & 58191.94 & 11980.90 & 169.90 $\pm$ 0.12 & 0.55 \\
  & 90401309004 & FPMA & 58198.04 & 2760.78 & 735.20 $\pm$ 0.52 & 0.49 \\
  &   & FPMB & 58198.04 & 2876.89 & 663.50 $\pm$ 0.48 & 0.48 \\
  & 90401309006 & FPMA & 58198.30 & 4539.90 & 761.20 $\pm$ 0.41 & 0.49 \\
  &   & FPMB & 58198.30 & 4762.39 & 681.20 $\pm$ 0.38 & 0.48 \\
  & 90401309008 & FPMA & 58201.53 & 3046.06 & 776.90 $\pm$ 0.51 & 0.48 \\
  &   & FPMB & 58201.53 & 3214.09 & 689.60 $\pm$ 0.47 & 0.47 \\
  & 90401309010 & FPMA & 58201.86 & 2660.36 & 782.70 $\pm$ 0.54 & 0.48 \\
  &   & FPMB & 58201.86 & 2801.44 & 697.00 $\pm$ 0.50 & 0.47 \\
  & 90401309012 & FPMA & 58212.20 & 12334.25 & 695.40 $\pm$ 0.24 & 0.47 \\
  &   & FPMB & 58212.20 & 12964.31 & 624.30 $\pm$ 0.22 & 0.46 \\
  & 90401309013 & FPMA & 58224.95 & 1834.38 & 672.60 $\pm$ 0.61 & 0.46 \\
  &   & FPMB & 58224.95 & 1934.06 & 597.70 $\pm$ 0.59 & 0.45 \\
  & 90401309014 & FPMA & 58225.29 & 9208.80 & 677.90 $\pm$ 0.27 & 0.46 \\
  &   & FPMB & 58225.29 & 9708.55 & 603.80 $\pm$ 0.25 & 0.45 \\
  & 90401309016 & FPMA & 58241.80 & 13792.23 & 572.90 $\pm$ 0.21 & 0.43 \\
  &   & FPMB & 58241.80 & 14431.53 & 514.60 $\pm$ 0.21 & 0.42 \\
  & 90401309018 & FPMA & 58255.16 & 2733.45 & 504.50 $\pm$ 0.48 & 0.42 \\
  &   & FPMB & 58255.16 & 2879.63 & 457.10 $\pm$ 0.40 & 0.41 \\
  & 90401309019 & FPMA & 58255.63 & 9444.10 & 503.40 $\pm$ 0.25 & 0.42 \\
  &   & FPMB & 58255.63 & 9938.34 & 455.80 $\pm$ 0.21 & 0.41 \\
  & 90401309021 & FPMA & 58297.17 & 21867.83 & 301.10 $\pm$ 0.12 & 0.43 \\
  &   & FPMB & 58297.17 & 22852.97 & 272.20 $\pm$ 0.11 & 0.42 \\
  & 90401309033 & FPMA & 58388.92 & 24892.60 & 124.20 $\pm$ 0.07 & 0.31 \\
  &   & FPMB & 58388.92 & 25342.94 & 116.50 $\pm$ 0.07 & 0.31 \\
  & 90401309035 & FPMA & 58397.30 & 18569.88 & 53.27 $\pm$ 0.05 & 0.44 \\
  &   & FPMB & 58397.30 & 18640.73 & 50.71 $\pm$ 0.05 & 0.44 \\
  & 90401309037 & FPMA & 58404.95 & 37682.99 & 14.04 $\pm$ 0.02 & 0.42 \\
  &   & FPMB & 58404.95 & 37608.34 & 13.18 $\pm$ 0.02 & 0.41 \\
  & 90401324002 & FPMA & 58259.25 & 6570.79 & 486.70 $\pm$ 0.27 & 0.42 \\
  &   & FPMB & 58259.25 & 6931.40 & 438.00 $\pm$ 0.25 & 0.41 \\
  & 90501311002 & FPMA & 58567.83 & 28666.65 & 13.82 $\pm$ 0.02 & 0.43 \\
  &   & FPMB & 58567.83 & 28590.56 & 12.92 $\pm$ 0.02 & 0.42 \\
  & 90501337002 & FPMA & 58721.31 & 44577.57 & 14.25 $\pm$ 0.02 & 0.43 \\
  &   & FPMB & 58721.31 & 44311.46 & 13.28 $\pm$ 0.02 & 0.43 \\
\hline               
Swift J1753.5-0127 & 30001148002 & FPMA & 56913.42 & 40434.75 & 12.32 $\pm$ 0.02 & 0.42 \\
  &   & FPMB & 56913.42 & 40476.89 & 11.69 $\pm$ 0.02 & 0.41 \\
\hline
\end{longtable}

\clearpage

\pagestyle{empty}
\begin{longrotatetable}

\tiny

\begin{longtable}{@{\extracolsep{0pt}}lcccccccccccccccc}

\caption{Best-fitting Results of the \emph{NuSTAR} sample}\label{tab:res}\\
%
\hline
  ObsID &$T_{\rm{in}}$ (keV) &$N_{\rm{disk}}$ &$q$ &$i$ (deg.) & $R_{\rm{in}}$ ($R_{\rm{ISCO}}$) &$\Gamma$ & log${\rm{\xi}}$ & $A_{\rm{Fe}}$ & $kT_{\rm{e}}$ (keV) &$R_{\rm{f}}$ &$N_{\rm{rel}}$ ($\times 10^{-4}$) & $N_{\rm{xil}}$ ($\times 10^{-4}$) &$\chi^2/\nu$\\
 \hline   
 \endfirsthead
\caption{continued.}\\
\hline\hline
  ObsID &$T_{\rm{in}}$ (keV) &$N_{\rm{disk}}$ &$q$ &$i$ (deg.) & $R_{\rm{in}}$ ($R_{\rm{g}}$) &$\Gamma$ & log${\rm{\xi}}$ & $A_{\rm{Fe}}$ & $kT_{\rm{e}}$ (keV) &$R_{\rm{f}}$  &$N_{\rm{rel}}$ ($\times 10^{-4}$) & $N_{\rm{xil}}$ ($\times 10^{-4}$) &$\chi^2/\nu$\\
 \hline
\endhead
\endfoot
\multicolumn{17}{c}{GRS 1716-249$^{\rm (a)}$}\\ 
80201034006 & 0.45$^{+0.01}_{-0.01}$ & 2609.8$^{+404.3}_{-397.3}$ & 2.5$^{+0.3}_{-0.3}$ & 9.8$^{+0.9}_{-1.6}$ & 4.6$^{+0.8}_{-1.0}$ & 1.584$^{+0.007}_{-0.006}$ & 3.91$^{+0.07}_{-0.10}$ & 2.8$^{+0.2}_{-0.3}$ & 48.5$^{+4.9}_{-4.7}$ & 0.16$^{+0.02}_{-0.02}$ & 267.4$^{+10.2}_{-10.3}$ & - & \\ 
80201034007 & 0.46$^{+0.01}_{-0.01}$ & 2927.5$_{-551.2}^{+416.7}$ & 10.0* & ... & 7.9$^{+0.5}_{-0.7}$ & 1.515$^{+0.005}_{-0.006}$ & 4.06$^{+0.02}_{-0.03}$ & ... & 339.4$^{+36.1}_{-35.7}$ & 0.39$^{+0.04}_{-0.03}$ & 292.1$^{+10.8}_{-8.2}$ & -  & $\frac{5696.9}{5600}$\\ 
90202055002 & 0.49$^{+0.01}_{-0.01}$ & 2050.2$_{-242.9}^{+244.1}$ & 3.3$^{+0.4}_{-0.6}$ & ... & 8.3$_{-1.1}^{+1.4}$ & 1.663$^{+0.007}_{-0.007}$ & 4.11$^{+0.03}_{-0.03}$ & ... & 400* & 0.48$^{+0.04}_{-0.05}$ & 191.9$_{-8.2}^{+9.8}$ & -  & \\ 
90202055004 & 0.51$^{+0.01}_{-0.01}$ & 1634.4$_{-149.5}^{+229.4}$ & 10.0* & ... & 63.7$_{-16.0}^{+25.3}$ & 1.771$^{+0.006}_{-0.007}$ & 3.94$^{+0.07}_{-0.06}$ & ... & 87.7$^{+17.7}_{-14.3}$ & 0.21$^{+0.04}_{-0.03}$ & 217.2$_{-12.2}^{+7.7}$ & - & \\ 
\hline 

\multicolumn{17}{c}{GRS 1739-278$^{\rm (b)}$}\\ 
80101050002  & - & - & - & - & - & 1.824$^{+0.035}_{-0.034}$ & - & - & 11.3$^{+4.0}_{-2.5}$ & -  & 44.2$^{+3.8}_{-3.5}$ & - \\
80102101002  & - & - & - & - & - & 1.68$^{+0.013}_{-0.008}$ & - & - & $>$ 34.4 & - & 265.5$^{+6.9}_{-5.4}$ & - & $\frac{2733.27}{2845}$\\
80102101004  & - & - & - & - & - & 1.699$^{+0.036}_{-0.019}$ & - & - & $>$ 30.4 & - & 82.6$^{+4.7}_{-3.7}$ & - \\
80102101005  & - & - & - & - & - & 1.670$^{+0.016}_{-0.014}$ & - & - & $>$ 175.6 & - & 166.8$^{+5.2}_{-6.3}$ & - \\
\hline
\multicolumn{17}{c}{GRS 1915+105$^{\rm (c)}$}\\ 
10002004001 & 0.35$^{+0.01}_{-0.00}$ & 184770.9$^{+24440.3}_{-32299.6}$ & 2.2$^{+0.1}_{-0.1}$ & $60^\dag$ & 2.8$^{+0.5}_{-0.2}$ & 1.942$^{+0.003}_{-0.003}$ & 4.50* & 3.7$^{+0.1}_{-0.1}$ & 16.6$^{+0.3}_{-0.3}$ & 2.54$^{+0.12}_{-0.14}$ & 141.1$^{+5.8}_{-4.3}$ & 30.9$^{+3.9}_{-3.7}$ & \\ 
80401312002 & 0.48$^{+0.01}_{-0.01}$ & 6891.3$^{+1061.8}_{-1013.5}$ & 3.8$^{+0.7}_{-0.4}$ & ... & 8.4$^{+1.2}_{-1.2}$ & 1.952$^{+0.006}_{-0.005}$ & 4.01$^{+0.03}_{-0.02}$ & ... & 65.7$^{+9.2}_{-7.7}$ & 1.27$^{+0.11}_{-0.08}$ & 108.0$^{+4.9}_{-4.6}$ & 15.0$^{+1.9}_{-2.3}$ & $\frac{3938.3}{3602}$\\ 
90202045002 & 0.40$^{+0.01}_{-0.01}$ & 53310.0$^{+9836.9}_{-8274.3}$ & 2.4$^{+0.2}_{-0.1}$ & ... & 3.3$^{+1.1}_{-1.0}$ & 1.970$^{+0.003}_{-0.003}$ & 4.40$^{+0.02}_{-0.02}$ & ... & 35.2$^{+1.7}_{-1.6}$ & 2.76$^{+0.09}_{-0.06}$ & 127.6$^{+2.4}_{-3.1}$ & 19.7$^{+3.3}_{-2.6}$ & \\ 
\hline 
\multicolumn{17}{c}{GS 1354-64$^{\rm (d)}$}\\ 
90101006002 & - & - & 4.4$^{+3.2}_{-2.2}$ & 29.0$^{+3.7}_{-3.3}$ & 56.9$^{+48.4}_{-25.8}$ & 1.505$^{+0.010}_{-0.011}$ & 3.03$^{+0.24}_{-0.19}$ & 0.5* & 46.1$^{+11.9}_{-8.4}$ & 0.14$^{+0.03}_{-0.04}$ & 20.3$^{+1.8}_{-1.2}$ & - & \\ 
90101006004 & - & - & 2.7$^{+1.2}_{-0.4}$ & ... & 9.8$^{+11.9}_{-8.3}$ & 1.623$^{+0.007}_{-0.006}$ & 2.97$^{+0.07}_{-0.06}$ & ... & 18.7$^{+1.0}_{-0.5}$ & 0.24$^{+0.08}_{-0.03}$ & 83.6$^{+1.4}_{-1.8}$ & - & $\frac{3871.9}{3470}$\\ 
90101006006 & - & - & 4.6$^{+2.9}_{-1.5}$ & ... & 44.1$^{+37.4}_{-16.5}$ & 1.663$^{+0.003}_{-0.003}$ & 3.48$^{+0.06}_{-0.07}$ & ... & 19.1$^{+0.6}_{-0.6}$ & 0.25$^{+0.04}_{-0.04}$ & 82.4$^{+2.6}_{-2.7}$ & - & \\ 
\hline 
\multicolumn{17}{c}{GX 339-4$^{\rm (d)}$}\\ 
80001013002 & - & - & 2.48$^{+0.05}_{-0.04}$ & $40^{\dag}$ & 1.18$^{+0.00}_{-0.00}$ & 1.585$^{+0.002}_{-0.002}$ & 1.93$^{+0.03}_{-0.03}$ & 5.3$^{+0.1}_{-0.1}$ & 240.8$^{+2.1}_{-3.1}$ & 0.255$^{+0.009}_{-0.005}$ & 36.6$^{+0.3}_{-0.5}$ & -  & \\ 
80001013004 & - & - & 2.85$^{+0.05}_{-0.05}$ & ... & 1.0* & 1.588$^{+0.002}_{-0.003}$ & 3.01$^{+0.04}_{-0.05}$ & ... & 316.9$^{+4.9}_{-4.5}$ & 0.275$^{+0.008}_{-0.006}$ & 59.0$^{+1.1}_{-0.9}$ & -  & \\ 
80001013006 & - & - & 6.25$^{+0.04}_{-0.05}$ & ... & 4.13$^{+0.13}_{-0.13}$ & 1.591$^{+0.003}_{-0.004}$ & 2.97$^{+0.07}_{-0.03}$ & ... & 400* & 0.272$^{+0.005}_{-0.005}$ & 94.9$^{+0.6}_{-1.3}$ & -  & \\ 
80001013008 & - & - & 2.83$^{+0.06}_{-0.07}$ & ... & 1.0* & 1.579$^{+0.006}_{-0.004}$ & 3.09$^{+0.03}_{-0.07}$ & ... & 400* & 0.310$^{+0.005}_{-0.005}$ & 125.0$^{+1.7}_{-0.8}$ & - & \\ 
80001013010 & - & - & 2.04$^{+0.08}_{-0.10}$ & ... & 31.43$^{+1.19}_{-0.65}$ & 1.598$^{+0.004}_{-0.003}$ & 1.73$^{+0.02}_{-0.03}$ & ... & 67.9$^{+2.0}_{-1.4}$ & 0.127$^{+0.001}_{-0.002}$ & 18.2$^{+0.3}_{-0.1}$ & - & \\ 
80001015001 & - & - & 2.87$^{+0.03}_{-0.03}$ & ... & 1.0* & 1.752$^{+0.004}_{-0.003}$ & 3.44$^{+0.01}_{-0.01}$ & ... & 48.8$^{+1.0}_{-2.2}$ & 0.311$^{+0.004}_{-0.006}$ & 198.7$^{+1.0}_{-2.4}$ & - & \\ 
80102011002 & - & - & 7.46$^{+0.05}_{-0.10}$ & ... & 3.47$^{+0.06}_{-0.04}$ & 1.748$^{+0.003}_{-0.003}$ & 3.51$^{+0.02}_{-0.04}$ & ... & 400* & 0.395$^{+0.015}_{-0.007}$ & 38.3$^{+0.6}_{-0.4}$ & - & $\frac{16674.3}{15205}$\\ 
80102011004 & - & - & 5.07$^{+0.04}_{-0.07}$ & ... & 2.88$^{+0.05}_{-0.06}$ & 1.669$^{+0.002}_{-0.003}$ & 3.30$^{+0.05}_{-0.06}$ & ... & 400* & 0.254$^{+0.004}_{-0.005}$ & 43.2$^{+0.4}_{-0.3}$ & - & \\ 
80102011006 & - & - & 10.0* & ... & 3.63$^{+0.05}_{-0.07}$ & 1.656$^{+0.004}_{-0.004}$ & 2.99$^{+0.02}_{-0.02}$ & ... & 400* & 0.239$^{+0.003}_{-0.012}$ & 41.1$^{+0.5}_{-0.9}$ & - & \\ 
80102011010 & - & - & 2.33$^{+0.05}_{-0.05}$ & ... & 1.34$^{+0.00}_{-0.01}$ & 1.640$^{+0.002}_{-0.003}$ & 3.00$^{+0.07}_{-0.02}$ & ... & 400* & 0.165$^{+0.003}_{-0.002}$ & 24.7$^{+0.6}_{-0.2}$ & - & \\ 
80102011012 & - & - & 2.66$^{+0.04}_{-0.03}$ & ... & 11.65$^{+0.37}_{-0.30}$ & 1.657$^{+0.002}_{-0.002}$ & 2.62$^{+0.03}_{-0.03}$ & ... & 187.2$^{+9.4}_{-2.9}$ & 0.169$^{+0.003}_{-0.005}$ & 11.9$^{+0.2}_{-0.1}$ & - & \\ 
80302304002 & - & - & 1.73$^{+0.05}_{-0.04}$ & ... & 1.0* & 1.648$^{+0.002}_{-0.003}$ & 2.93$^{+0.05}_{-0.02}$ & ... & 47.6$^{+1.0}_{-1.0}$ & 0.165$^{+0.004}_{-0.003}$ & 5.7$^{+0.1}_{-0.1}$ & - & \\ 
80302304004 & - & - & 10.0* & ... & 3.33$^{+0.04}_{-0.05}$ & 1.564$^{+0.003}_{-0.003}$ & 3.16$^{+0.02}_{-0.04}$ & ... & 400* & 0.318$^{+0.005}_{-0.009}$ & 81.0$^{+2.4}_{-0.7}$ & - & \\ 
80302304005 & - & - & 10.0* & ... & 2.95$^{+0.17}_{-0.08}$ & 1.612$^{+0.002}_{-0.009}$ & 3.02$^{+0.02}_{-0.01}$ & ... & 400* & 0.286$^{+0.006}_{-0.004}$ & 66.4$^{+0.5}_{-0.5}$ & - & \\ 
80502325008 & - & - & 2.80$^{+0.04}_{-0.05}$ & ... & 1.0* & 1.681$^{+0.003}_{-0.003}$ & 3.18$^{+0.03}_{-0.04}$ & ... & 400* & 0.256$^{+0.006}_{-0.005}$ & 28.9$^{+0.4}_{-0.2}$ & - & \\ 
90401369004 & - & - & 0.0* & ... & 1.0* & 1.581$^{+0.001}_{-0.002}$ & 1.91$^{+0.12}_{-0.09}$ & ... & 90.3$^{+3.1}_{-2.1}$ & 0.160$^{+0.003}_{-0.005}$ & 43.0$^{+0.7}_{-0.7}$ & - & \\

\hline 
\multicolumn{17}{c}{H1743-322$^{\rm (d)}$}\\ 
80001044002 & - & - & 2.7$_{-0.6}^{+1.5}$ & 51.9$_{-5.6}^{+2.0}$ & 4.4$_{-0.6}^{+1.2}$ & 1.642$_{-0.007}^{+0.006}$ & 1.8$_{-0.1}^{+0.1}$ & 0.5* & 150.1$_{-9.7}^{+4.9}$ & 0.34$_{-0.10}^{+0.03}$ & 76.0$_{-1.3}^{+1.6}$ & - & \\ 
80001044004 & - & - & 2.3$_{-0.2}^{+0.3}$ & ... & 4.2$_{-0.7}^{+0.5}$ & 1.690$_{-0.006}^{+0.006}$ & 2.0$_{-0.1}^{+0.1}$ & ... & 88.9$_{-12.5}^{+7.3}$ & 0.38$_{-0.06}^{+0.03}$ & 85.2$_{-1.2}^{+1.3}$ & - & \\ 
80001044006 & - & - & 1.9$_{-0.3}^{+0.2}$ & ... & 8.3$_{-3.1}^{+0.7}$ & 1.690$_{-0.011}^{+0.009}$ & 1.4$_{-0.2}^{+0.5}$ & ... & 76.8$_{-7.0}^{+6.8}$ & 0.43$_{-0.06}^{+0.03}$ & 73.8$_{-1.6}^{+1.6}$ & - & \\ 
80002040002 & - & - & 2.2$_{-0.4}^{+0.3}$ & ... & 6.9$_{-1.2}^{+0.8}$ & 1.630$_{-0.008}^{+0.012}$ & 0.1* & ... & 290.0$_{-18.5}^{+19.9}$ & 0.29$_{-0.04}^{+0.05}$ & 57.2$_{-1.9}^{+1.5}$ & - & $\frac{12438.6}{12080}$\\ 
80202012002 & - & - & 1.1$_{-0.3}^{+0.6}$ & ... & 2.6$_{-1.0}^{+1.4}$ & 1.848$_{-0.009}^{+0.026}$ & 1.7$_{-0.0}^{+0.2}$ & ... & 66.9$_{-3.5}^{+7.6}$ & 0.42$_{-0.04}^{+0.13}$ & 80.4$_{-2.9}^{+0.7}$ & -  & \\ 
80202012004 & - & - & 2.3$_{-0.3}^{+0.7}$ & ... & 3.8$_{-0.6}^{+1.2}$ & 1.864$_{-0.007}^{+0.027}$ & 1.5$_{-0.1}^{+0.1}$ & ... & 158.4$_{-5.4}^{+5.2}$ & 0.42$_{-0.04}^{+0.11}$ & 81.5$_{-2.9}^{+0.8}$ & - & \\ 
80202012006 & - & - & 2.5$_{-0.1}^{+0.2}$ & ... & 3.9$_{-0.4}^{+0.5}$ & 1.661$_{-0.007}^{+0.006}$ & 0.1* & ... & 178.5$_{-25.8}^{+94.4}$ & 0.27$_{-0.04}^{+0.03}$ & 57.5$_{-1.4}^{+1.9}$ & - & \\ 
90401335002 & - & - & 1.9$_{-0.2}^{+0.1}$ & ... & 6.5$_{-1.7}^{+0.6}$ & 1.704$_{-0.020}^{+0.010}$ & 1.7$_{-0.1}^{+0.1}$ & ... & 72.0$_{-2.9}^{+5.3}$ & 0.31$_{-0.15}^{+0.05}$ & 78.4$_{-1.9}^{+5.2}$ & - & \\ 
\hline 
\multicolumn{17}{c}{IGR J17091-3624$^{\rm (d)}$}\\ 
80001041002 & - & - & 2.2$^{+0.5}_{-0.3}$ & 52.3$^{+7.3}_{-10.0}$ & 7.6$^{+7.6}_{-4.9}$ & 1.739$^{+0.009}_{-0.009}$ & 1.3$^{+0.6}_{-0.5}$ & 0.5* & 400* & 0.56$^{+0.12}_{-0.11}$ & 39.0$^{+0.9}_{-0.9}$ & - & \\ 
80202014002 & - & - & 2.7$^{+1.8}_{-0.5}$ & ... & 13.2$^{+20.1}_{-9.6}$ & 1.737$^{+0.014}_{-0.013}$ & 2.5$^{+0.1}_{-0.1}$ & ... & 54.7$^{+12.4}_{-8.6}$ & 0.40$^{+0.18}_{-0.10}$ & 42.1$^{+1.1}_{-1.2}$ & - & $\frac{3128.1}{3603}$\\ 
80202014004 & - & - & 1.8$^{+0.5}_{-0.7}$ & ... & 10.1$^{+10.0}_{-6.2}$ & 1.740$^{+0.009}_{-0.008}$ & 2.6$^{+0.1}_{-0.1}$ & ... & 36.7$^{+5.4}_{-3.7}$ & 0.26$^{+0.05}_{-0.05}$ & 45.3$^{+0.8}_{-0.8}$ & - & \\ 
\hline 
\multicolumn{17}{c}{MAXI J1348-630$^{\rm (a)}$}\\ 
80402315002 & 0.50$^{+0.01}_{-0.01}$ & 16290.3$^{+1737.6}_{-2093.8}$ & 2.2$^{+0.2}_{-0.1}$ & 17.3$^{+1.1}_{-1.5}$ & 4.0$^{+0.8}_{-0.4}$ & 1.687$^{+0.003}_{-0.003}$ & 3.58$^{+0.03}_{-0.03}$ & 2.1$^{+0.1}_{-0.1}$ & 25.5$^{+1.3}_{-0.8}$ & 0.18$^{+0.02}_{-0.01}$ & 1930.9$^{+19.3}_{-17.7}$ & - & \\ 
80402315004 & 0.53$^{+0.03}_{-0.01}$ & 9955.0$^{+639.8}_{-1094.3}$ & 10.0* & ... & 16.1$^{+2.5}_{-3.5}$ & 1.688$^{+0.004}_{-0.006}$ & 3.45$^{+0.05}_{-0.03}$ & ... & 23.3$^{+1.3}_{-1.2}$ & 0.17$^{+0.02}_{-0.01}$ & 1952.1$^{+20.9}_{-28.4}$ & - & \\ 
80502304002 & 0.58$^{+0.02}_{-0.01}$ & 744.8$^{+136.4}_{-60.1}$ & 2.6$^{+0.1}_{-0.1}$ & ... & 3.5$^{+0.5}_{-0.2}$ & 1.644$^{+0.018}_{-0.005}$ & 3.43$^{+0.06}_{-0.23}$ & ... & 89.6$^{+11.7}_{-9.4}$ & 0.25$^{+0.01}_{-0.03}$ & 444.5$^{+30.9}_{-16.7}$ & - & $\frac{7810.4}{7131}$\\ 
80502304004 & 0.54$^{+0.01}_{-0.01}$ & 1209.0$^{+52.2}_{-145.3}$ & 3.1$^{+0.5}_{-0.2}$ & ... & 7.8$^{+0.6}_{-2.0}$ & 1.639$^{+0.008}_{-0.004}$ & 3.21$^{+0.04}_{-0.09}$ & ... & 80.3$^{+17.8}_{-14.3}$ & 0.16$^{+0.01}_{-0.01}$ & 419.7$^{+15.7}_{-11.7}$ & - & \\ 
80502304006 & 0.59$^{+0.01}_{-0.02}$ & 335.3$^{+96.6}_{-58.7}$ & 2.4$^{+0.1}_{-0.1}$ & ... & 5.9$^{+1.9}_{-1.2}$ & 1.637$^{+0.005}_{-0.010}$ & 3.09$^{+0.07}_{-0.05}$ & ... & 89.5$^{+3.1}_{-3.6}$ & 0.16$^{+0.03}_{-0.01}$ & 287.5$^{+3.9}_{-4.8}$ & - & \\ 
\hline 
\multicolumn{17}{c}{MAXI J1535-571$^{\rm (c)}$}\\ 
90301013002 & 0.40$^{+0.01}_{-0.01}$ & 217048.3$^{+50274.2}_{-36802.9}$ & 5.9$^{+3.1}_{-3.6}$ & 56.4$^{+6.7}_{-2.5}$ & 3.3$^{+0.7}_{-0.9}$ & 1.83$^{+0.04}_{-0.04}$ & 3.7$^{+0.2}_{-0.3}$ & 1.2$^{+0.8}_{-0.5}$ & 19.7$^{+1.7}_{-0.7}$ & 0.77$^{+0.15}_{-0.13}$ & 863.5$^{+107.2}_{-122.6}$ & 111.9$^{+53.7}_{-33.3}$ & $\frac{1747.4}{1671}$\\ 
\hline 
\multicolumn{17}{c}{MAXI J1813-095$^{\rm (d)}$}\\ 
80402303002 & - & - & 3.7$^{+3.0}_{-1.1}$ & 17.9$^{+5.9}_{-6.9}$ & 7.5$^{+6.2}_{-3.7}$ & 1.61$^{+0.01}_{-0.02}$ & 3.4$^{+0.3}_{-0.2}$ & 1.5$^{+0.4}_{-0.3}$ & 400* & 0.15$^{+0.06}_{-0.03}$ & 49.1$^{+1.8}_{-3.2}$ & - & \\ 
80402303004 & - & - & 3.9$^{+2.3}_{-1.1}$ & ... & 7.4$^{+6.9}_{-3.2}$ & 1.65$^{+0.01}_{-0.01}$ & 2.8$^{+0.2}_{-0.2}$ & ... & 307.2$^{+59.7}_{-76.8}$ & 0.12$^{+0.03}_{-0.02}$ & 41.2$^{+1.2}_{-1.4}$ & - & $\frac{2812.0}{3322}$\\ 
80402303006 & - & - & 6.2$^{+2.6}_{-2.4}$ & ... & 14.9$^{+6.0}_{-4.5}$ & 1.69$^{+0.01}_{-0.01}$ & 3.1$^{+0.1}_{-0.1}$ & ... & 400* & 0.15$^{+0.03}_{-0.02}$ & 39.4$^{+0.7}_{-0.9}$ & - & \\ 
\hline 
\multicolumn{17}{c}{MAXI J1820+070$^{\rm (c)}$}\\ 
90401309002 & 1.5* & 4.6$^{+0.1}_{-0.1}$ & 10.0* & $63^{\dag}$ & 207.7$^{+4.0}_{-3.3}$ & 1.510$^{+0.001}_{-0.001}$ & 0.1* & 2.47$^{+0.04}_{-0.04}$ & 40.6$^{+0.6}_{-0.5}$ & 0.13$^{+0.00}_{-0.00}$ & 525.7$^{+2.9}_{-2.9}$  & 33.2$^{+1.0}_{-1.2}$ & \\

90401309004 & 1.5* & 26.0$^{+0.8}_{-0.9}$ & 2.0$^{+0.1}_{-0.1}$ & ... & 2.9$^{+0.0}_{-0.0}$ & 1.640$^{+0.002}_{-0.002}$ & 1.98$^{+0.02}_{-0.01}$ & ... & 39.7$^{+0.9}_{-0.9}$ & 0.54$^{+0.03}_{-0.01}$ & 1628.1$^{+12.2}_{-14.7}$ & 0.1* & \\

90401309006 & 1.21$^{+0.01}_{-0.01}$ & 84.1$^{+2.4}_{-2.1}$ & 2.0$^{+0.1}_{-0.1}$ & ... & 12.7$^{+0.5}_{-0.3}$ & 1.577$^{+0.001}_{-0.001}$ & 3.20$^{+0.02}_{-0.01}$ & ... & 24.4$^{+0.6}_{-0.7}$ & 0.21$^{+0.01}_{-0.01}$ & 1536.1$^{+12.9}_{-13.7}$ & 180.1$^{+4.8}_{-4.7}$ & \\

90401309008 & 0.73$^{+0.01}_{-0.01}$ & 598.2$^{+12.5}_{-12.1}$ & 1.7$^{+0.1}_{-0.1}$ & ... & 19.1$^{+0.3}_{-0.4}$ & 1.591$^{+0.002}_{-0.003}$ & 3.80$^{+0.02}_{-0.04}$ & ... & 43.3$^{+1.3}_{-1.2}$ & 0.81$^{+0.05}_{-0.03}$ & 1256.3$^{+15.7}_{-14.8}$ & 300.0$^{+6.2}_{-5.9}$ & \\

90401309010 & 0.96$^{+0.02}_{-0.01}$ & 147.5$^{+3.6}_{-4.5}$ & 2.1$^{+0.1}_{-0.1}$ & ... & 25.1$^{+0.6}_{-0.7}$ & 1.567$^{+0.002}_{-0.002}$ & 3.90$^{+0.01}_{-0.01}$ & ... & 42.5$^{+1.1}_{-1.4}$ & 1.32$^{+0.02}_{-0.02}$ & 996.9$^{+11.0}_{-12.9}$ & 318.4$^{+9.5}_{-8.6}$ & \\

90401309012 & 1.08$^{+0.01}_{-0.01}$ & 113.8$^{+4.6}_{-3.4}$ & 2.4$^{+0.1}_{-0.1}$ & ... & 17.3$^{+0.5}_{-0.5}$ & 1.595$^{+0.002}_{-0.003}$ & 3.25$^{+0.01}_{-0.01}$ & ... & 29.7$^{+0.7}_{-0.6}$ & 0.23$^{+0.00}_{-0.01}$ & 1464.5$^{+9.9}_{-12.7}$ & 144.1$^{+1.8}_{-1.7}$ & $\frac{21638.5}{19059}$\\

90401309013 & 0.63$^{+0.01}_{-0.01}$ & 875.6$^{+31.2}_{-24.5}$ & 2.0$^{+0.0}_{-0.0}$ & ... & 3.6$^{+0.1}_{-0.1}$ & 1.728$^{+0.003}_{-0.002}$ & 1.74$^{+0.04}_{-0.02}$ & ... & 79.7$^{+3.6}_{-2.0}$ & 0.73$^{+0.02}_{-0.02}$ & 1555.8$^{+10.4}_{-9.8}$ & 0.7$^{+0.0}_{-0.0}$ & \\

90401309014 & 1.03$^{+0.01}_{-0.01}$ & 138.6$^{+5.2}_{-3.8}$ & 2.8$^{+0.1}_{-0.1}$ & ... & 18.2$^{+0.6}_{-0.9}$ & 1.623$^{+0.001}_{-0.001}$ & 3.24$^{+0.01}_{-0.02}$ & ... & 31.6$^{+0.8}_{-1.0}$ & 0.26$^{+0.01}_{-0.01}$ & 1383.7$^{+10.2}_{-11.7}$ & 137.5$^{+4.8}_{-5.2}$ & \\

90401309016 & 0.53$^{+0.01}_{-0.01}$ & 2974.1$^{+107.1}_{-117.7}$ & 10.0* & ... & 27.0$^{+1.0}_{-1.1}$ & 1.651$^{+0.003}_{-0.002}$ & 3.73$^{+0.02}_{-0.01}$ & ... & 133.1$^{+3.1}_{-3.0}$ & 0.84$^{+0.02}_{-0.02}$ & 1029.4$^{+10.4}_{-15.7}$ & 169.6$^{+3.6}_{-5.2}$ & \\

90401309018 & 0.87$^{+0.03}_{-0.02}$ & 182.5$^{+7.9}_{-6.1}$ & 2.2$^{+0.1}_{-0.1}$ & ... & 11.1$^{+0.4}_{-0.4}$ & 1.677$^{+0.002}_{-0.002}$ & 3.35$^{+0.02}_{-0.02}$ & ... & 53.7$^{+1.7}_{-1.9}$ & 0.40$^{+0.01}_{-0.01}$ & 1034.1$^{+11.1}_{-10.9}$ & 62.8$^{+1.9}_{-1.8}$ & \\

90401309019 & 0.73$^{+0.01}_{-0.01}$ & 462.7$^{+17.1}_{-12.1}$ & 10.0* & ... & 22.1$^{+0.8}_{-0.8}$ & 1.678$^{+0.001}_{-0.001}$ & 3.51$^{+0.02}_{-0.01}$ & ... & 96.0$^{+3.4}_{-4.3}$ & 0.58$^{+0.02}_{-0.02}$ & 991.9$^{+10.6}_{-7.5}$ & 135.9$^{+3.4}_{-2.9}$ & \\

90401309021 & 0.74$^{+0.01}_{-0.01}$ & 287.8$^{+5.1}_{-4.8}$ & 2.6$^{+0.2}_{-0.1}$ & ... & 12.5$^{+0.4}_{-0.5}$ & 1.649$^{+0.002}_{-0.001}$ & 3.23$^{+0.01}_{-0.01}$ & ... & 262.7$^{+4.6}_{-4.8}$ & 0.37$^{+0.02}_{-0.01}$ & 843.9$^{+6.0}_{-9.1}$ & 54.2$^{+1.3}_{-1.6}$ & \\

90401309033 & 0.70$^{+0.01}_{-0.01}$ & 406.4$^{+35.0}_{-38.5}$ & 3.2$^{+0.3}_{-0.2}$ & ... & 8.1$^{+0.9}_{-0.6}$ & 1.880$^{+0.003}_{-0.003}$ & 3.33$^{+0.03}_{-0.02}$ & ... & 400* & 0.78$^{+0.05}_{-0.04}$ & 203.9$^{+3.4}_{-4.7}$ & 10.3$^{+2.0}_{-2.1}$ & \\

90401309035 & 1.32$^{+0.05}_{-0.03}$ & 1.9$^{+0.3}_{-0.2}$ & 10.0* & ... & 52.8$^{+9.3}_{-6.8}$ & 1.660$^{+0.005}_{-0.006}$ & 2.90$^{+0.07}_{-0.07}$ & ... & 228.4$^{+24.2}_{-17.7}$ & 0.15$^{+0.01}_{-0.02}$ & 156.0$^{+2.0}_{-1.9}$ & 3.5$^{+0.7}_{-0.5}$ & \\

90401309037 & - & - & 10.0* & ... & 63.3$^{+9.8}_{-10.8}$ & 1.730$^{+0.005}_{-0.006}$ & 2.91$^{+0.08}_{-0.10}$ & ... & 400* & 0.16$^{+0.02}_{-0.02}$ & 38.1$^{+0.4}_{-0.4}$ & 2.1$^{+0.3}_{-0.2}$ & $\frac{7878.0}{7307}$\\

90401324002 & 0.80$^{+0.02}_{-0.02}$ & 271.6$^{+36.1}_{-29.8}$ & 10.0* & ... & 21.7$^{+2.8}_{-1.8}$ & 1.685$^{+0.004}_{-0.004}$ & 3.46$^{+0.05}_{-0.04}$ & ... & 82.5$^{+10.4}_{-9.1}$ & 0.49$^{+0.04}_{-0.04}$ & 986.5$^{+29.6}_{-23.5}$ & 112.6$^{+10.4}_{-11.0}$ & \\

90501311002 & - & - & 10.0* & ... & 112.4$^{+17.9}_{-20.6}$ & 1.723$^{+0.005}_{-0.005}$ & 2.85$^{+0.10}_{-0.10}$ & ... & 80.0$^{+11.6}_{-12.6}$ & 0.12$^{+0.01}_{-0.02}$ & 32.8$^{+0.6}_{-0.9}$ & 3.1$^{+0.6}_{-0.4}$ & \\

90501337002 & - & - & 10.0* & - & 400.0* & 1.706$^{+0.004}_{-0.005}$ & 2.71$^{+0.06}_{-0.09}$ & ... & 133.6$^{+17.3}_{-15.3}$ & 0.15$^{+0.02}_{-0.02}$ & 37.5$^{+0.6}_{-0.6}$ & 3.0$^{+0.4}_{-0.6}$ & \\ 
\hline 
\multicolumn{17}{c}{Swift J1753.5-0127$^{\rm (b)}$}\\ 
30001148002 & - & - & - & - & - & 1.736$^{+0.004}_{-0.004}$ & - & - & $>$ 55.8 & - & - & - & $\frac{1069.5}{987}$\\ 
\hline

\hline

\end{longtable}

\footnotesize{

Note:\\
    - The letter `a' indicates the spectral model {\tt {TBabs*(diskbb+relxillCp)}} adopted in the sources.\\
    - The letter `b' indicates the spectral model {\tt {TBabs*nthcomp}} adopted in the sources.\\
    - The letter `c' indicates the spectral model {\tt {TBabs*(diskbb+relxillCp+xillverCp)}} adopted in the sources.\\
    - The letter `d' indicates the spectral model {\tt {TBabs*relxillCp}} adopted in the sources.\\
    - The symbol `$*$' indicates the parameters are pegged at lower/upper limit adopted in the spectral fits.\\
    - The symbol `$\dag$' indicates the parameters are fixed at the values given in Table 1.\\
    - The symbol `...' indicates the parameters are linked in the joint fits for a given source.\\
    - The symbol `-' for $T_{\rm in}$ and $N_{\rm disk}$ in the {\tt diskbb} model indicates that the disk component is not required in the fits for a given observation. $T_{\rm in}$ = 0.05\,keV and $N_{\rm disk}=10^{-5}$ are fixed in the joint fits. 
}

\end{longrotatetable}

\clearpage

\begin{appendix} 
\section{Comparison with the existing constraints}

\subsection{GRS 1716-249}
The spectral analysis for the four epochs, i.e., 80201034006, 80201034007, 90202055002, and 90202055004, was also done by \cite{jiang2020}. In their spectral fits, the power-law model {\tt cutoffpl} is used to account for the coronal emission, and the {\tt relconvlp*reflionx} is adopted to model the relativistic disk reflection, where {\tt kerrconv} is a relativistic blurring function based on ray-tracing simulations \citep{brenneman2006} and {\tt reflionx} accounts for the reflection from an ionized accretion disk of constant density \citep{ross2005}. 
Their results show that the inclination angle is low with $i \sim 30^{\circ}$, and the accretion disk is of low iron abundance and highly ionized, which are consistent with our results. The Comptonization continua are harder with the photon index in the range of 1.3-1.5 than the ones in our fitting results, and the reflection fraction is about 0.25. Note that, the modeling of the relativistic disk reflection with {\tt relconvlp*reflionx} is based on the assumption of a simple lamppost geometry, and more importantly, the empirical reflection fraction in {\tt relconvlp*reflionx} is different from the physical reflection fraction in the reflection model {\tt relxillCp} adopted in our work \citep{dauser2016}. 

\subsection{GRS 1739-278}

The four \emph{NuSTAR} spectra of GRS 1739-278 studied in this work are well described by a power-law continuum with an exponential cutoff, which agrees with the results in previous studies \citep{furst2016,yan2020}. Joint fits of \emph{XMM-Newton} and \emph{NuSTAR} for one epoch observation on MJD 57281 (ObsID: 80101050002) by \cite{furst2016} revealed the signatures of reflection. Their spectral fits indicate a low reflection fraction with $R = 0.043$ for a hard spectrum with photon index $\Gamma = 1.39$, which agrees well with the low end of the $R$-$\Gamma$ correlation. 



\subsection{GRS 1915+105}



The \emph{NuSTAR} spectrum of GRS 1915+105 in 2012 (ObsID: 10002004001) was analyzed in previous studies with the reflection model \citep{mil2013,zhanga2019,zhangb2019,shreeram2020}. In \cite{mil2013}, the spectral model mainly consists of ({\tt kerrconv}$\times${\tt reflionx})+{\tt cutoffpl}. Assuming the disk inner edge being at ISCO, their best fits prefer a high BH spin with $a=0.98$ and a large inclination angle for this binary system with $i=72^{\circ}$, and strongly exclude the fitting results with a much high cutoff energy. As a comparison, in our fits of this observation with the assumptions of $a=0.998$ and $i=60^{\circ}$, the fitted inner radius of the disk is about 2 $R_{\rm ISCO}$, and the electron temperature $kT_{\rm e}$ is about 17\,keV. However, the derived photon index in our fits is about 1.9, which is larger than the one ($\sim 1.7$) in \cite{mil2013}. The same observation was also studied using the relativistic reflection models {\tt relxill\_NK} \citep{bambi2017} which belongs to the RELXILL family \citep{zhanga2019,zhangb2019,shreeram2020}. The fitted photon index in those studies is large with $\Gamma = 2.13$, instead.


\subsection{GS 1354-64}

The three \emph{NuSTAR} spectra of GS 1354-64 were analyzed by \cite{yan2020} in which the Comptonization model ({\tt nthcomp}) is adopted, while the disk reflection is not included, for simplicity. In their spectral fits, the photon index of the Comptonization continuum is $\Gamma = 1.49, 1.57$, 1.65, and the fitted electron temperature is low with $kT_{\rm e} \sim$ 10--20\,keV, which are consistent with the results in this work.

Moreover, two of the three \emph{NuSTAR} spectra (ObsID: 90101006002, 90101006004), were also analyzed in previous studies with the reflection model \citep{ei2016,xu2018,tripathi2021}.
\cite{ei2016} found a large truncation radius for the accretion disk with $R_{\rm in} = 700_{-500}^{+200} R_{\rm ISCO}$ for 90101006002, but a small truncation radius, which is close to $R_{\rm ISCO}$ for 90101006004. In this work, we also find the similar results that the disk is truncated at a large radius with $R_{\rm in} \simeq 44_{-17}^{+37} R_{\rm ISCO}$ for 90101006002, but at a small radius $R_{\rm in} = 9.8_{-8.3}^{+11.8} R_{\rm ISCO}$ for 90101006004. And a low iron abundance with $A_{\rm Fe} \sim 0.5$ is preferred in both \cite{ei2016} and our work.


\subsection{GX 339-4}

The \emph{NuSTAR} spectra of GX 339-4 studied in this work mainly consist of the observations in 2013 \citep{furst2015,wj2018,jiang2019}, 2015 \citep{wj2018,jiang2019}, and 2017 \citep{garcia2019,wangj2020}. For the observations in 2013 (ObsID: 80001013002-04-06-08-10) and 2015 (ObsID: 80102011002-04-06-10-12), the reflection models in the {\tt relxill} family are implemented by \cite{wj2018} in the joint fits to \emph{NuSTAR} and \emph{Swift} data. In their results, the inclination angle is constrained with $i \sim 40^{\circ}$, which supports the assumption of the inclination angle in our fits. And the electron temperatures are found to be large with $kT_{\rm e} \sim$ hundreds of kiloelectronvolts and the disk inner radii are close to the ISCO, which are consistent with our fitting results. The constrained photon index is in the range of 1.5-1.7, as we find in this work. And both \cite{wj2018} and our work found that the spectra prefers to a high iron abundance. Note that, such a high iron overabundance may be due to the assumed constant electron density of the accretion disk $n_{\rm e} = 10^{15} ~ \rm cm^{-3}$ \citep{jiang2019}.





\subsection{H1743-322}

The \emph{NuSTAR} spectra of H1743-322 were also analyzed with the phenomenological models \citep{yan2020,wangpj2022} and the reflection models in the {\tt relxill} family \citep{ingram2016,chand2020,stiele2021}. A low iron abundance with $A_{\rm Fe} \sim 0.5$ and a small disk truncation radius of a few times $R_{\rm ISCO}$ are suggested in the spectral fits \citep{ingram2016,stiele2021}. And the fitted photon index and the reflection fraction by \cite{stiele2021} follow the derived correlation in this work. 


\subsection{IGR J17091-3624}

The \emph{NuSTAR} spectra of IGR J17091-3624 were analyzed by \cite{xu2017} with the reflection models in the {\tt relxill} family. In their spectral results, the constrained inner radii are about of tens of  gravitational radii, and the iron abundance is low with $A_{\rm Fe} \sim 0.7$, which are consistent with the results in this work. And the fitted photon index and the reflection fraction are slightly lower than the ones in our results.


\subsection{MAXI J1348-630}

Two of the \emph{NuSTAR} spectra of MAXI J1348-630 studied in this work (ObsIDs: 80402315002 and 80402315004) were also analyzed in \cite{sudip2021} and \cite{jia2022} with the reflection models in the {\tt relxill} family. In their results, the disk temperature at the inner radius is about 0.5\,keV. For the Comptonizing source, the photon index $\Gamma \sim 1.7$, and the electron temperature $kT_{\rm e} \sim 30$ keV. These measurements are consistent with the ones in this work. We notice that the disk in their fits is truncated at $\sim 10R_{\rm g}$, while the inner radius in our results is about $\sim 5R_{\rm g}$. Furthermore, their reflection fraction is slightly smaller than the one in our fits.

\subsection{MAXI J1535-571}

The \emph{NuSTAR} spectra of MAXI J1348-630 were analyzed by \cite{xuyj2018} and \cite{dong2022}. The observation (ObsID = 90301013002) was fitted by \cite{xuyj2018} with different flavors of the reflection models in the {\tt relxill} family. Our fitting results, in terms of the disk properties (the inner temperature $kT_{\rm bb}$, iron abundance, and the ionization parameter), the Comptonization spectrum (the electron temperature $kT_{\rm e}$ and the photon index), and the reflection fraction, are consistent with the ones in \cite{xuyj2018}.  


\subsection{MAXI J1813-095}

The three \emph{NuSTAR} spectra of MAXI J1813-095 were stacked by \cite{jiang2022} 
for the spectral analysis with the reflection models in the {\tt relxill} family. 
In their results, the inclination of the source is found 
to be low with $i \sim 22^{\circ}$ which supports our conclusion 
of a low inclination angle of $i \sim 18^{\circ}$. And, the disk properties (the inner temperature $T_{\rm in}$, iron abundance, and the ionization parameter), the Comptonization spectrum (the electron temperature $kT_{\rm e}$ and the photon index) in our fits are consistent with the ones in \cite{jiang2022}.  


\subsection{MAXI J1820+070}

The \emph{NuSTAR} spectra of MAXI J1820+070 in the 2018 outburst were analyzed by \cite{buisson2019} with a dual-lamppost
model (two lampposts at different coronal heights) {\tt relxilllpCp} in the {\tt relxill} family. In their results, the inclination angle is fitted to be $\sim 30^{\circ}$, while it is fixed with $i =63^{\circ}$ \citep{atr2020} in this work. Note that there is a large misalignment ($>40^{\circ}$) between the inclination angle of the jet and the binary orbit in MAXI J1820+070 \citep{pou2022}.


\subsection{Swift J1753.5-0127}

The \emph{NuSTAR} spectra of Swift J1753.5-0127 studied in this work are well-described by a power-law continuum with an exponential cutoff. 
The Comptonization spectrum, in terms of the electron temperature $kT_{\rm e}$ and the photon index, in our fits are consistent with the results in previous studies \citep{yan2020}, which include the \emph{Swift} data as well.

\end{appendix}

\end{document}